\documentclass[10pt]{article}
\usepackage[margin=.8in,left=.8in]{geometry}

\usepackage{amsthm}
\usepackage{amssymb}
\usepackage{amsmath}
\usepackage{mathtools}
\usepackage{adjustbox}

\usepackage[table]{xcolor}

\usepackage{stmaryrd}    
\usepackage{xfrac}
\usepackage{enumitem}\setlist{nosep}

\usepackage[safe]{tipa} 

\usepackage{hhline}

\usepackage{tensor}

\usepackage{tikz}
\usetikzlibrary{
  backgrounds, 
  decorations.pathmorphing, 
  decorations.markings,
  decorations.text,shapes.geometric, arrows.meta, positioning, calc
}

\usepackage{mathrsfs} 
\DeclareMathAlphabet{\mathpzc}{OT1}{pzc}{m}{it} 

\usepackage{hyperref}
\hypersetup{
  colorlinks = true,
  linkcolor= darkblue, citecolor=darkblue            
}

\hypersetup{
        colorlinks=true,
        linkcolor=darkblue,
        filecolor=darkblue,      
        urlcolor=black,
        citecolor=darkblue,
        linktoc=page
    }

\usepackage{cleveref}

\definecolor{darkblue}{rgb}{0.05,0.25,0.65}
\definecolor{darkgreen}{RGB}{20,140,10}
\definecolor{lightgray}{rgb}{0.9,0.9,0.9}
\definecolor{darkorange}{RGB}{200,100,5}
\definecolor{darkyellow}{rgb}{.91,.91,0}
\definecolor{orangeii}{RGB}{200,100,5}
\definecolor{lightblue}{RGB}{243, 250, 255}

\newtheorem{theorem}{Theorem}[section]

\newtheorem{lemma}[theorem]{Lemma}

\newtheorem{proposition}[theorem]{Proposition}
\newtheorem{corollary}[theorem]{Corollary}
\theoremstyle{definition}
\newtheorem{definition}[theorem]{Definition}

\newtheorem{example}[theorem]{Example}
\newtheorem{assumption}[theorem]{Assumption}

\newtheorem{remark}[theorem]{Remark}

\usepackage{accents}
\newlength{\dhatheight}

\let\PLAINthebibliography\thebibliography
\renewcommand\thebibliography[1]{
  \PLAINthebibliography{#1}
  \setlength{\parskip}{0.5pt}
  \setlength{\itemsep}{0.5pt plus .3ex}
}

\newcommand{\shape}{
  \raisebox{1pt}{\rm\normalfont\textesh}
}

\newcommand\bos[1]{\mathstrut\mkern2.5mu#1\mkern-14mu\raise1.7ex%
  \hbox{$\scriptstyle\rightsquigarrow$}}

\newcommand\bosonic[1]{\mathstrut\mkern2.5mu#1\mkern-14mu\raise1.7ex%
  \hbox{$\scriptstyle\rightsquigarrow$}}

\usepackage[cmtip,all]{xy}
\newcommand{\longsquiggly}{\xymatrix{{}\ar@{~>}[r]&{}}}

\usepackage{bbm} 

\usepackage[new]{old-arrows}   

\newcommand{\FR}{\mathbb{R}}

\newcommand{\dd}{\mathrm{d}}

\newcommand{\ddCE}
{\dd_{\mathrm{CE}}}

\newcommand{\CE}{\mathcal{E}}

\newcommand{\frg}
{\mathfrak{g}}

\newcommand{\Triv}
{\mathrm{Triv}}

\newcommand{\Null}
{\mathrm{Null}}

\newcommand{\DD}{\mathrm{D}}
\begin{document}

\setlength{\abovedisplayskip}{3pt}
\setlength{\belowdisplayskip}{3pt}
\setlength{\abovedisplayshortskip}{-3pt}
\setlength{\belowdisplayshortskip}{3pt}

\title{Higher Groupoids of Gauge Potentials \\ in Higher Gauge Theories \\ admitting Weights} 


\author{
  Grigorios Giotopoulos${}^{\ast}$
}

\maketitle

\thispagestyle{empty}

\begin{abstract}

We construct an explicit small model for the $\infty$-groupoid of
null-concordances of higher gauge curvatures by algorithmically
trivializing their nonlinear Bianchi identities.
More precisely, for a finite-type Sullivan minimal Bianchi algebra
equipped with positive integer weights, we organize weighted
potentials and higher gauge parameters at each fixed curvature
into an $\infty$-groupoid whose gauge transformations are linear
in the parameters.
 We prove that this model is naturally equivalent to the
$\infty$-groupoid of null-concordances through an intermediate
total-form model, with explicit projection and section maps
and simplicial homotopies.
 When nonempty, its connected components form a torsor for the
degree-zero cohomology of an associated weighted gauge complex,
and its higher homotopy groups are computed by the higher 
cohomology of this complex.

\end{abstract}

\vspace{.1cm}

\begin{center}
\begin{minipage}{13cm}
\tableofcontents
\end{minipage}
\end{center}

\medskip

\vfill

\hrule
\vspace{5pt}

{
\footnotesize
\noindent
\def\arraystretch{1}
\tabcolsep=0pt
\begin{tabular}{ll}
${}^*$\,
&
Mathematics, Division of Science; and
\\
&
Center for Quantum and Topological Systems,
\\
&
NYUAD Research Institute,
\\
&
New York University Abu Dhabi, UAE.  
\end{tabular}
\hfill
\href{https://ncatlab.org/nlab/show/Center+for+Quantum+and+Topological+Systems}{
\adjustbox{raise=-15pt}{
\includegraphics[width=3cm]{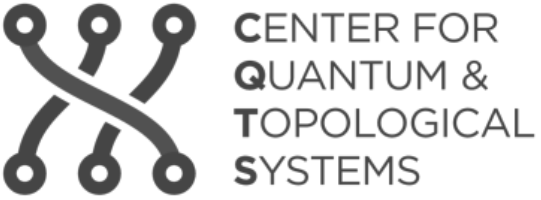}
}
}

\vspace{.2cm}

\noindent
The author acknowledges the support by {\it Tamkeen} under the 
{\it NYU Abu Dhabi Research Institute grant} {\tt CG008}.
}

\newpage

\medskip

\section{Introduction and Summary}
In \cite[\S~2.1.4]{GSS24}\cite[\S~4.1]{GSS-M5Brane}\cite[\S~4.1.1]{GSS26Survey} we have discussed a naturally defined groupoid of local gauge potentials for the supergravity C-field, obtained as a small explicit model of the 1-truncation of a canonically defined infinity-groupoid involving smooth concordances of duality-symmetric (``electromagnetic'') gauge potentials. Here we show how this construction of \cite{GSS24, GSS-M5Brane, GSS26Survey} extends beyond the 1-truncation, and beyond the special example of 11D supergravity, to the construction of a small (explicit, hands-on) infinity-groupoid of local gauge potentials for any higher gauge theory satisfying a mild condition (`admitting weights'). The latter generically holds for the higher gauge sectors of higher dimensional supergravity theories.

\paragraph{Bianchi identities and higher potentials.}
Nonlinear Bianchi identities govern the curvature forms of many higher gauge
theories, including the (duality-symmetric) flux sectors of supergravity \cite{GSS26Survey}. On (super-)manifold $M$,
writing these identities as 
\begin{align}\label{BianchiIdentitiesEquation}
\dd G^{(i)}=P^{(i)}(G)\, ,
\end{align}
their universal
consistency condition determines a differential graded commutative algebra
\[
 A\,=\,\FR_\dd [x^{(i)}]\big /\bigl(\dd x^{(i)}=P^{(i)}(x)\bigr)\,,
 \qquad \dd^2\,=\,0\, ,
\]
so that a solution is precisely a DGCA morphism $A\to\Omega^\bullet(M)$, sending
$x^{(i)}$ to $G^{(i)}$. In this paper we assume that $A$ is a Sullivan
minimal model, written $\mathrm{CE}(\mathfrak g)$, for a connected nilpotent
rational homotopy type, with scalars extended to $\mathbb R$ (Assumption~\ref{ass:main}, cf.~\cite{Sullivan1977}), so that $\mathfrak g$ is the corresponding dual $L_\infty$-algebra. The positive-weight hypothesis imposed below is an additional property of
the Bianchi algebra.

Locally on a contractible $U\hookrightarrow M$, these identities may be solved (i.e., trivialized) in terms of potentials $C$, which in turn
admit gauge transformations $\Lambda_1$ (again via trivialization, cf. Ex. \ref{OneExampleTrivializations}), further gauge-of-gauge transformations $\Lambda_2$ and so on. It follows that this object of local ``solutions'' does not form a bear set, since two gauge potentials $C$ and $C'$ are not said to be equivalent only if related by equality, but rather by an (invertible) \textit{gauge transformation} $\Lambda_1: C \rightarrow C'$, hence forming a ``groupoid''. But the same holds for the gauge transformations themselves, two of them are said to be equivalent if they are related by a gauge-of-gauge transformation. Since this pattern continues in principle to infinity, one expects this solution object to have the structure of a so called ``$\infty$-groupoid''. However, to organize these
data into an actual $\infty$-groupoid, one must further specify the compositions of gauge transformations and their higher
coherences -- which is not directly obvious when obtaining these by inductive trivializations without further specification. 

On the other hand, there exists a canonical, albeit more abstract, construction of a simplicial set directly
from the structure of the Bianchi algebra $\mathrm{CE}(\frg)$. This is the simplicial set of higher concordances,
whose $n$-simplices are
\[
 \begin{aligned}
 \bigl(\shape\Omega^1_{\mathrm{cl}}(U;\mathfrak g)\bigr)_n
 &\, :=\,
 \operatorname{Hom}_{\mathrm{DGCA}}\!\big(
 \mathrm{CE}(\mathfrak g),\Omega^\bullet(U\times\Delta^n)\big)\\
 &\,\cong\,
 \operatorname{MC}\!\big(\Omega^\bullet(U\times\Delta^n)\otimes\mathfrak g\big)\, .
 \end{aligned}
\]
Here faces and degeneracies are induced by pullback along simplex maps, and $\operatorname{MC}$ denotes Maurer--Cartan elements of total degree one of the corresponding $L_\infty$-algebra of forms valued in $\frg$. The symbol $\shape$ denotes the shape construction of Schreiber \cite[Def.~4.4.65]{dcct}\cite[Def.~9.1]{FSS23Character},
presented here as ``higher concordances'', as in \cite{GSS26Survey}.
For a fixed $U$, this is the smooth differential-form version of the simplicial
Maurer--Cartan construction employed in Hinich
\cite[Def.~2.1.1 and Prop.~2.2.3]{Hinich1997} and Getzler
\cite[\S4]{Getzler}, extending the earlier Deligne groupoid viewpoint
\cite{Deligne1986}.\footnote{Hinich and Getzler use polynomial forms on
parameter simplices. Here we use smooth total forms, up to the boundary,
as in the smooth integration approach of \cite{Henriques}. But crucially, we now have a further smooth parametrization over $U\hookrightarrow M$ as per Schreiber \cite{dcct}.} The underlying simplicial construction goes further back to Sullivan's
spatial realization of differential graded commutative algebras
\cite[\S8]{Sullivan1977}.
For our purposes, we are interested in the following variation: We fix a curvature $G$ and consider the corresponding
\emph{null}-concordances, with fixed endpoints
\[
\begin{aligned}
 \Null^\infty(G)(U)_n
 &\,=\,\left\{\widehat G\in
 \operatorname{Hom}_{\mathrm{DGCA}}\!\big(
 \mathrm{CE}(\mathfrak g),\Omega^\bullet(U\times\Delta^n\times I_t)\big)
 \;\middle|\;
 \begin{array}{l}
 \widehat G|_{t=0}=0,\\
 \widehat G|_{t=1}=\operatorname{pr}_U^*G
 \end{array}\right\}\\[4pt]
 &\,=\,\left\{\widehat G\in
 \operatorname{MC}\!\big(\Omega^\bullet(U\times\Delta^n\times I_t)\otimes\mathfrak g\big)
 \;\middle|\;
 \begin{array}{l}
 \widehat G|_{t=0}=0,\\
 \widehat G|_{t=1}=\operatorname{pr}_U^*G
 \end{array}\right\}.
\end{aligned}
\]

This is our Def.~\ref{HigherNullConc}, following \cite[\S~2.1.4]{GSS24}, which acts as a model for the space of morphisms between the curvatures $0$ and $G$ in $\shape\Omega^1_{\mathrm{cl}}(U;\mathfrak g)$. Its vertices are
null-concordances -- homotopies between 0 and the fixed curvature $G$, while its higher simplices are concordances between these
relative to the fixed curvature endpoints. The idea is that this object is supposed to universally encode potentials of $G$, their gauge transformations, and so on. The purpose of this paper is to construct an explicit \textit{linear
model} of trivializations (``solutions'' to the Bianchi identities) in terms of potentials and higher gauge parameters, under a
positive-weight hypothesis, and to prove its equivalence with the
$\infty$-groupoid of null-concordances at fixed curvature.

\begin{example}[\bf An example of C-field higher trivializations]\label{OneExampleTrivializations}
The characteristic algebra of the eleven-dimensional C-field is the minimal
model of the rational $4$-sphere \cite[\S~2.5]{Sati13}\cite[\S~4]{FSS15-M5WZW}\cite[Cor. 2.3]{FSS17},
\[
 \mathrm{CE}(\mathfrak l S^4)
 \,=\,
\FR_\dd [g_4, g_7]\big /  \bigl(\dd g_4=0,\ \dd g_7=\tfrac12 g_4^2\bigr)\, .
\]
Thus an $\mathfrak l S^4$-valued curvature on a (super-)manifold $M$ is a pair
of differential forms satisfying
\[
 \dd G_4\,=\,0,
 \qquad
 \dd G_7\,=\,\tfrac12 G_4G_4\, ,
\]
and hence, on a contractible chart $U$, successive applications of the Poincar\'e lemma
give potentials \cite[Eq. (6.7-6.11)]{CL94}
\[
 G_4\,=\,\dd C_3,
 \qquad
 G_7\,=\,\dd C_6+\tfrac12 C_3G_4\, .
\]

In the convention of \cite[Prop.~2.48]{GSS24} and
\cite[Def.~4.3]{GSS26Survey}, a gauge transformation
$(C_3,C_6)\to(C'_3,C'_6)$ is specified by forms $(B_2,B_5)$ with
\[
 C'_3-C_3\,=\,\dd B_2,
 \qquad
 C'_6-C_6\,=\,\dd B_5+\tfrac12 C'_3C_3\, .
\]
Note that this notion arises by \textit{choosing} to trivialize the difference of two potentials via 
$$
\dd(C_3'-C_3) \, = \, 0, \qquad \dd(C_6' - C_6 -\tfrac{1}{2} C_3' C_3) \, = \, 0\, ,
$$
but this choice is not unique (see Ex. \ref{WeightedCfieldTrivializations} below), and as this paper suggests, there are more convenient choices. The ``\textit{weighted}'' choice below has the additional advantage that its higher transformation equations and composition laws are linear in the potentials at fixed curvature. 

For two parallel such transformations, a gauge-of-gauge transformation $(B_2,B_5')\rightarrow (B_2', B_5')$ is specified by a
pair $(B_1,B_4)$ satisfying
\[
 B'_2-B_2\,=\,\dd B_1,
 \qquad B'_5-B_5\,=\,\dd B_4\, .
\]
In other words, it is a pair that trivializes the difference of two gauge transformations (between the same two potentials)
$$
\dd(B_2'-B_2) \, = \, 0, \quad \dd(B_5'-B_5) \, = \,0\, .
$$

Continuing the parallel-boundary rule gives parameter pairs of degrees
\[
 (2,5),\quad(1,4),\quad(0,3),\quad(-1,2),\quad(-2,1),\quad(-3,0)\, ,
\]
where negative-degree forms are zero. More explicitly, writing
$B^{[r]}=(B_{3-r},B_{6-r})$ for a level-$r$ parameter, the parallel-boundary
equations above extend to
\[
 (B^{[r-1]})' - B^{[r-1]}
 \,=\,\dd B^{[r]},\qquad 2\leq r\leq6.
\]
At each stage the two boundary transformations are parallel, i.e., they have the
same lower-dimensional source and target. 

These equations specify a continuation of the parallel-boundary rules, but
the cited low-level comparison does not by itself establish a full
higher groupoid with these rules. Namely, coherent composition is additional structure to be specified, and
already at level one the conventional composition contains products of
parameters (\cite[Lem.~4.4--4.5]{GSS26Survey}) -- which complicates the picture. 

But these equations specify \textit{only one way} of successive trivializations -- among several possibilities. The weighted formulation
below organizes the full hierarchy, including its coherences, by linear
equations.
\end{example}

\begin{example}[\bf Weighted higher C-field trivializations]\label{WeightedCfieldTrivializations}
Assigning weights $w(g_4)=1$ and $w(g_7)=2$ (equivalently, the supergravity
weights $3$ and $6$) to the generators, the weighted algorithm (Sec. \ref{WeightedBianchisSection}) gives the same equations for
$C_3,C_6$, but uses instead gauge parameters $(\Lambda_2,\Lambda_5)$ satisfying
\[
 C'_3-C_3\,=\,\dd\Lambda_2,
 \qquad
 C'_6-C_6\,=\,\dd\Lambda_5-\tfrac12\Lambda_2G_4\, .
\]
These map onto the preceding convention under the change of parameters
\[
 \Lambda_2\,=\,B_2,
 \qquad
 \Lambda_5\,=\,B_5+\tfrac12 B_2C_3\, ,
\]
since $\dd(B_2C_3)=\dd B_2\,C_3+B_2G_4$ and $C_3C_3=0$.
Continuing in the same fashion, the complete list of parallel-boundary equations is
\[
\begin{aligned}
 C'_3-C_3&\,=\,\dd\Lambda_2,
 &C'_6-C_6&\,=\,\dd\Lambda_5-\tfrac12\Lambda_2G_4,\\
 \Lambda'_2-\Lambda_2&\,=\,\dd\Lambda_1,
 &\Lambda'_5-\Lambda_5&\,=\,\dd\Lambda_4+\tfrac12\Lambda_1G_4,\\
 \Lambda'_1-\Lambda_1&\,=\,\dd\Lambda_0,
 &\Lambda'_4-\Lambda_4&\,=\,\dd\Lambda_3-\tfrac12\Lambda_0G_4,\\
 \Lambda'_0-\Lambda_0&\,=\,0,
 &\Lambda'_3-\Lambda_3&\,=\,\dd\Lambda_2^{[4]},\\
 &&(\Lambda_2^{[4]})'-\Lambda_2^{[4]}&\,=\,\dd\Lambda_1^{[5]},\\
 &&(\Lambda_1^{[5]})'-\Lambda_1^{[5]}&\,=\,\dd\Lambda_0^{[6]}\, .\footnote{The superscripts in the last three lines distinguish the remaining
$G_7$-sector parameters from the earlier $G_4$-sector parameters of the same
form degree.}
\end{aligned}
\]
 A quick glance at these equations shows that they are organized by the ``weighted differential'' 
\[
 \DD_r^G(\alpha_{3-r},\beta_{6-r})
 \,:=\,
 \bigl(\dd\alpha_{3-r},\,
 \dd\beta_{6-r}+\tfrac{(-1)^r}{2}\alpha_{3-r}G_4\bigr)\,. \footnote{For parallel
higher transformations, the preceding unweighted equations are converted
to these equations by
$\beta_{6-r}=B_{6-r}+\tfrac12B_{3-r}C_3$ for $1\leq r\leq3$, and by
$\beta_{6-r}=B_{6-r}$ for $r\geq4$, with the common source potential $C_3$
held fixed.} 
\]
In particular, the (globally defined) curvature-dependent terms do not disappear
at the higher levels.

Crucially, by the linearity of these equations, composition of weighted parameters with compatible boundaries is simply \textit{addition}, and
the identities 
$$
\DD_r^G\DD_{r+1}^G=0
$$
organize these equations and all their
simplicial compatibility conditions. In general, the weighted algorithm
therefore yields an $\infty$-groupoid -- that of
$\Triv_W^\infty(G)(M)$ from Definition~\ref{WGPC:def:parameters}. The notation stands here for ``weighted trivializations'' of $G$ on $M$. 
Its defining linearity means that, whenever it is nonempty and an origin
$C_*$ is chosen, it is completely encoded by the weighted gauge cochain
complex $L^\bullet$ of Definition~\ref{WGPC:def:gaugecomplex}, through the
Dold--Kan construction (Remark~\ref{WGPC:rem:doldkan}). 
\end{example}

\paragraph{Positive weights and the range of examples.}
The general input in our discussion is a Sullivan minimal DGCA
$$ 
A=\mathrm{CE}(\mathfrak g)\, =\, \big(\FR[x^{(i)}],\,\dd_{\mathrm{CE}}\big),\qquad
\dd_{\mathrm{CE}}x^{(i)}=P^{(i)}(x) \, ,
$$ whose generators carry positive
integer weights $w_i$ such that every monomial in $P^{(i)}$ has weight
$w_i$. Note that the differential preserves this additional grading, and weights need
not equal differential-form degrees. Our precise hypotheses are stated in
Assumption~\ref{ass:main}, and a criterion for a positive weight assignment
is given in Lemma~\ref{PositiveWeightCriterion}.

This hypothesis covers the supergravity Bianchi systems motivating
this work, including the C-field and the descendant systems, as 
it is, in particular, stable under the (rational) cyclification and toroidification constructions (Rem. \ref{SufficientConditions}, Ex. \ref{4SphereWeightsViaMinkowski}). Moreover, it is compatible with the scaling weights of the
supergravity fluxes and the corresponding scale invariance of the theories (cf. Cor. \ref{ScaleInvarianceCorollary}), which provides evidence for the broad physical scope of
the algorithm. For any curvature $G$ in this class the construction is explicit:
Potentials solve 
$$
\DD_0^G C=G\, ,
$$ and similarly for their (higher) gauge transformations, e.g.,
$$
C-\hat{C} \, = \, \DD_1^G(\Lambda_1), \qquad \Lambda_1-\hat{\Lambda}_1 \, = \, \DD_2^G(\Lambda_2)
$$
and so on, where here the differential is defined by \[\DD_r^G\,:=\,\dd+(-1)^rR(-,G)
,
 \qquad
  R^{(i)}(C,G)
 \,:=\,\frac1{w_i}\sum_j w_j C^{(j)}
 \frac{\partial P^{(i)}}{\partial x^{(j)}}(G) \, .
\]

\paragraph{The comparison theorem and its computational consequence.}

We construct this explicit linear model for null-concordances as weighted trivializations by passing through the intermediate
$\infty$-groupoid of weighted total-form trivializations
(Def.~\ref{WGPC:def:forms}). Its simplices are potentials on
$M\times\Delta^n$ satisfying one total differential-form equation, and integration
over faces and Whitney reconstruction identify it, up to simplicial
homotopy, with the original weighted trivializations model
(Theorem~\ref{WGPC:prop:comparison}). Independently, the weighted projection
and section exhibit it as a strong simplicial deformation retract of the
null-concordance model of Def.~\ref{HigherNullConc}
(Theorem~\ref{WNCH:thm:equivalence}). Consequently,
\[
 \boxed{\quad
 \Null^\infty(G)(M)
 \,\simeq\,
 \Triv_{W,\mathrm{total}}^\infty(G)(M)
 \,\simeq\,
 \Triv_W^\infty(G)(M)\, .
 \quad}
\]
Note these are homotopy equivalences of Kan simplicial sets, natural under
pullback in $M$, rather than strict isomorphisms of simplicial sets.
Combining the two theorems gives Corollary~\ref{WNCH:cor:gaugehomotopy}:
a global null-concordance exists exactly when a global weighted
trivialization exists, and, after choosing a null-concordance $\widehat G_*$,
\[
 \pi_r\bigl(\Null^\infty(G)(M),\widehat G_*\bigr)
 \,\cong\,
 H^{-r}(L^\bullet)
 \,=\,
 \frac{\ker\DD_r^G}{\operatorname{im}\DD_{r+1}^G},
 \qquad r\geq1.
\]
The identification uses the homotopy--homology theorem for the Dold--Kan
construction \cite[Cor.~8.3]{Kan1958}. When nonempty, the set of components
is naturally a torsor for
\[
 H^0(L^\bullet)
 \,=\,\frac{\ker\DD_0^G}{\operatorname{im}\DD_1^G}\, ,
\]
while choosing $C_*=P_W(\widehat G_*)$ identifies that torsor with this vector
space by $[\widehat G]\mapsto[P_W(\widehat G)-C_*]$.

\paragraph{Relation to the C-field comparison.}
For $\mathfrak g=\mathfrak lS^4$, the comparison of
\cite[Prop.~2.48]{GSS24} is developed in
\cite[Lem.~4.5 and Prop.~4.6]{GSS26Survey} into an equivalence of
\emph{1-truncated} local groupoids. Namely, arrows there are (gauge-of-gauge)
equivalence classes of gauge transformations. For the fixed-curvature null-concordance situation, our result extends
this comparison in three directions:
\begin{enumerate}
\item It retains every higher morphism and coherence, giving an equivalence
of the full $\infty$-groupoids. In particular, it recovers the agreement
of equivalence classes while identifying all higher automorphism groups.
\item The linear weighted model computes these groups explicitly as the
cohomology of the gauge complex.
\item The construction works for every weighted Bianchi algebra satisfying
Assumption~\ref{ass:main}, extending beyond the local
C-field example.
\end{enumerate}
The extension to the full higher structure
and the cohomological calculation are the additional conclusions, while agreement
at the 1-truncated level is already present in the cited work for the case of $\frg = \mathfrak{l}S^4$.

\paragraph{Null-concordances and globally completed gauge fields.}
The importance of higher concordances goes beyond a particular choice of
potential formulas. A curvature satisfying nonlinear Bianchi identities
determines a cocycle in nonabelian de Rham cohomology. A global completion
specifies a non-rational charge cocycle together with a homotopy identifying
its character with this curvature cocycle. This is the differential
nonabelian-cohomological description developed in
\cite{FSS23Character,SS25Flux} and applied to (superspace) supergravity in
\cite{GSS24, GSS-M5Brane, GSIIA, GSS26Survey}.

Briefly, for a classifying space $\mathcal A$\footnote{In the relevant higher smooth context, this means the constant smooth $\infty$-stack over $\mathrm{CartSp}$ associated to its homotopy type.} with compatible real
homotopy type $\mathcal A_{\mathbb R}$, the corresponding differential cohomology classifying object
has the form\footnote{Here $\Omega^1_{\mathrm{cl}}(-;\mathfrak g)$ is the sheaf over $\mathrm{CartSp}$ that assigns to any Cartesian space the set of curvatures solving the Bianchi identities, i.e., $\Omega^1_{\mathrm{cl}}(U;\mathfrak g)\,\cong \, \mathrm{Hom}_{\mathrm{DGCA}}\big(\mathrm{CE}(\frg),\, \Omega^\bullet(U)\big)$.}
\[
 \mathcal A_{\mathrm{diff}}
 \,=\,
 \Omega^1_{\mathrm{cl}}(-;\mathfrak g)
 \mathop{\times}\limits^h_{\mathcal A_{\mathbb R}}\mathcal A.
\]
In the higher smooth setting of the cited sources (see also \cite{dcct, Schreiber24, Gi25}), the real classifying
object is presented by the shape of the curvature presheaf,
\[
 \mathcal A_{\mathbb R}
 \,\simeq\,\shape\Omega^1_{\mathrm{cl}}(-;\mathfrak g)\, .
\]
A globally completed field is then encoded by a morphism from $M$ into $\mathcal{A}_\mathrm{diff}$. In other words, it represents triples $(G,\chi, \widehat{C})$ that fit in the following
square, where $\widehat{C}$ is the displayed comparison homotopy
\begin{align}\label{DiffCohomologyDiagram}
\begin{tikzpicture}[baseline=(current bounding box.center)]
 \node (m) at (0,1.6) {$M$};
 \node (a) at (7,1.6) {$\mathcal A$};
 \node (f) at (0,0) {$\Omega^1_{\mathrm{cl}}(-;\mathfrak g)$};
 \node (r) at (7,0) {$\displaystyle\shape\Omega^1_{\mathrm{cl}}(-;\mathfrak g)$\, . };
 \draw[->] (m) -- node[above] {$\chi$} (a);
 \draw[->] (m) -- node[left] {$G$} (f);
 \draw[->] (a) -- node[right] {$\mathrm{ch}_{\mathcal A}$} (r);
 \draw[->] (f) -- node[below] {$\eta^{\shape}$} (r);
 \node at (3.5,0.8) {$\widehat{C}\;\Downarrow$};
\end{tikzpicture}
\end{align}
Here $\eta^{\shape}$ is the canonical map including each curvature
as a vertex of its integrating $\infty$-groupoid, while
the right-hand arrow is the nonabelian character map
(\cite[Def.~9.3]{FSS23Character}, \cite[Eq.~(246)]{GSS26Survey}).
For a fixed curvature $G$, a global completion field therefore consists of a charge cocycle
$\chi:M\to\mathcal A$ and a specified comparison homotopy -- where the latter \textit{ought to} encode all the associated local gauge potentials of $G$ and their induced (higher) compatibility relations due to the relation with $\chi$.

In particular, when the charge
cocycle is the zero cocycle, or is equipped with a chosen trivialization,
the comparison homotopy gives a null-homotopy
from the zero cocycle to the curvature cocycle.
The defining square then specializes to
\[
\begin{tikzpicture}[baseline=(current bounding box.center)]
 \node (m) at (0,1.6) {$M$};
 \node (z) at (7,1.6) {$*$};
 \node (f) at (0,0) {$\Omega^1_{\mathrm{cl}}(-;\mathfrak g)$};
 \node (r) at (7,0) {$\displaystyle\shape\Omega^1_{\mathrm{cl}}(-;\mathfrak g)$};
 \draw[->] (m) -- (z);
 \draw[->] (m) -- node[left] {$G$} (f);
 \draw[->] (z) -- node[right] {$0$} (r);
 \draw[->] (f) -- node[below] {$\eta^{\shape}$} (r);
 \node at (3.5,0.8) {$\widehat G\;\Downarrow$};
\end{tikzpicture}
\]
With the curvature \(G\) and the zero-charge map fixed, the comparison homotopy ought to encode the (now global) potential data and admits a differential-form null-concordance representative. Indeed, the space of these choices, together with their gauge transformations and higher transformations, is modeled by \(\Null^\infty(G)(M)\) (Def.~\ref{HigherNullConc}), which is in turn equivalent (Cor.~\ref{WNCH:cor:gaugehomotopy}) to the $\infty$-groupoid of (weighted) trivializations $\mathrm{Triv}_W^\infty(G)(M)$(Def.~\ref{WGPC:def:forms}).

\paragraph{Outlook: \v{C}ech representatives of globally completed fields.}
For an arbitrary globally completed field $(\chi,G,\widehat{C})$, the charge
need not be globally trivial. Assuming $\mathcal A$ is pointed and connected,
we may choose on a (differentiably) good cover $\{U_a\}$ local charge trivializations (homotopies)
$\tau_a:*\Rightarrow\chi|_{U_a}$, which exist because, for a contractible chart $U_a$,
$\operatorname{Map}(U_a,\mathcal A)\simeq\mathcal A$,
and $\mathcal A$ is connected. Having oriented the comparison homotopy as
$\widehat{C}:\mathrm{ch}_{\mathcal A}(\chi)\Rightarrow\eta^{\shape}(G)$,
composition gives local null-homotopies
\[
 n_a\,:=\,\widehat{C}|_{U_a}\circ\mathrm{ch}_{\mathcal A}(\tau_a)
 \;:\;0\Longrightarrow\eta^{\shape}(G|_{U_a})\, ,
\]
and after choosing differential-form representatives, our comparison
identifies their local null-concordance $\infty$-groupoids with those
of weighted potentials and higher gauge parameters
(Theorems~\ref{WNCH:thm:equivalence} and~\ref{WGPC:prop:comparison}).
This is the local mechanism underlying the role of null-concordances
in \cite[Def.~2.46, Prop.~2.48]{GSS24} and
\cite[\S4.1, Ex.~4.7]{GSS26Survey} \cite[Prop 1.1]{GSIIA}.

On $U_{ab}$, the restrictions of $\tau_a$ and $\tau_b$ are paths in
$\operatorname{Map}(U_{ab},\mathcal A)$ from the constant zero-charge
map to $\chi|_{U_{ab}}$. Their concatenated difference is thus the based loop (at the constant map)
\[
\ell_{ab}\,:=\,\tau_b^{-1}\circ\tau_a\, ,
\]
which yields over $U_{ab}$,
\[
\begin{aligned}
n_b\circ\mathrm{ch}_{\mathcal A}(\ell_{ab})
\, &=\, 
\bigl(\widehat{C}\circ\mathrm{ch}_{\mathcal A}(\tau_b)\bigr)
\circ\mathrm{ch}_{\mathcal A}(\tau_b^{-1}\circ\tau_a)
\\
&\simeq \, \widehat{C}
\circ\mathrm{ch}_{\mathcal A}(\tau_b\circ \tau_b^{-1}\circ\tau_a)
\\
&\simeq \, 
\widehat{C}\circ\mathrm{ch}_{\mathcal A}(\tau_a)
\\
&=:\, n_a.
\end{aligned}
\]

Thus local potentials obtained from a globally completed field remain
related via homotopy, up to its charge transitions. A full \v{C}ech presentation
must retain these transitions and their higher compatibility data,
which are already implicit in the global field $(\chi,G,\widehat{C})$.


\medskip


The local comparison underlying the \v{C}ech description in
\cite[Ex.~4.7]{GSS26Survey} is established at the level of potentials
and gauge transformations modulo gauge-of-gauge transformations.
Our equivalence with the full null-concordance $\infty$-groupoid supplies
the corresponding higher local comparison. This forms the basis
for constructing \v{C}ech data from weighted potentials and higher gauge
parameters, together with the corresponding locally constant charge
data on higher overlaps.

Briefly, a chosen filling of the character loop
$\mathrm{ch}_{\mathcal A}(\ell_{ab})$ translates, through the
differential-form presentation and our comparison, into a weighted
gauge transformation between the local potentials. Compatible higher
fillings similarly give higher weighted gauge parameters.
Writing $C_a$ for the local weighted potentials, the first expected
matching equations are
\[
 \begin{aligned}
 C_b-C_a
 &\,=\,\DD_1^G\Lambda_1^{ab}
 &&\text{on }U_{ab},\\
 \Lambda_1^{ac}-\Lambda_1^{ab}-\Lambda_1^{bc}
 &\,=\,\DD_2^G\Lambda_2^{abc}
 &&\text{on }U_{abc},
 \end{aligned}
\]
with all forms restricted to the indicated intersections. Notice the second
equation expresses the agreement of the direct transition with the
additive composite up to a higher weighted gauge transformation.
For $\mathfrak g=\mathfrak lS^4$, these equations read
\[
 \begin{aligned}
 C_3^b-C_3^a&\,=\,\dd\Lambda_2^{ab},\\
 C_6^b-C_6^a&\,=\,\dd\Lambda_5^{ab}-\tfrac12\Lambda_2^{ab}G_4,\\
 \Lambda_2^{ac}-\Lambda_2^{ab}-\Lambda_2^{bc}
 &\,=\,\dd\Lambda_1^{abc},\\
 \Lambda_5^{ac}-\Lambda_5^{ab}-\Lambda_5^{bc}
 &\,=\,\dd\Lambda_4^{abc}+\tfrac12\Lambda_1^{abc}G_4\,.
 \end{aligned}
\]

When a higher charge filling fails to exist, its homotopy class should
instead be retained as locally constant transition data, subject to
further compatibility conditions. Note that this only obstructs continuing the chosen
trivializations, rather than the existence of the completed field.
For $S^4$, its $3$-connectedness allows the first stages to be filled, with
the first potentially nonzero charge class occurring on fivefold overlaps
in $\pi_4(S^4)\cong\mathbb Z$. Explicitly, the next $G_4$-sector
parameter is a function $\Lambda_0^{abcd}$ on quadruple overlaps, with
\[
 \Lambda_1^{bcd}-\Lambda_1^{acd}
 +\Lambda_1^{abd}-\Lambda_1^{abc}
 \,=\,\dd\Lambda_0^{abcd}.
\]
The expected primary charge equation on fivefold overlaps is
\[
 \Lambda_0^{bcde}-\Lambda_0^{acde}+\Lambda_0^{abde}
 -\Lambda_0^{abce}+\Lambda_0^{abcd}
 \,=\,-q^{abcde},
 \qquad \dd q^{abcde}\,=\,0,
\]
where $q$ is a locally constant \v{C}ech $4$-cocycle, $\delta q=0$,
representing the primary charge. With integral normalization and a
suitable charge representative, $q$ is integer-valued (the minus sign
follows the displayed descent convention). Local form descent alone
gives a real locally constant cocycle, while matching it to the prescribed
integral charge representative is part of the expected construction.

The real form hierarchy also suggests coupled secondary data on
eightfold overlaps. Note, however, that these expected fivefold and eightfold matching
conditions are \textit{not independent equations} describing two separate
abelian higher gerbes, as the parameter hierarchies are already coupled
on double and triple overlaps by the terms
$-\tfrac12\Lambda_2^{ab}G_4$ and
$+\tfrac12\Lambda_1^{abc}G_4$ displayed above. For the globally completed ``component'' corresponding to $G_7$, this deduction is also readily visible at the global level of curvatures, since it is not closed. Accordingly, the
secondary matching condition must incorporate this coupling to the
primary charge data.

Establishing this explicit descent description and its compatibility with the
character map remains a further task for the case of $S^4$-completed fields, and yet more generally  for any weighted Sullivan model
$\mathrm{CE}(\mathfrak g)$ satisfying our hypotheses with a chosen
non-rational completion classifying space $\mathcal A$ of that real homotopy type.

\medskip

\paragraph{\bf Acknowledgements.} The author thanks Urs Schreiber for valuable comments on a late version of the manuscript. The author acknowledges considerable use of the LLM model GPT-6 Astra, especially for certain technical steps in the proofs of the theorems and for connecting to the relevant literature. Every step has been independently verified by the author. 

\medskip

\section{The $\infty$-groupoid of weighted trivializations}
\paragraph{\bf Weighted Bianchi identities.}\label{WeightedBianchisSection}

Fix some\footnote{That is, to include the cases where several different forms of the same degree appear. If there's a single form in each degree, then the ordering is of course unique.} ordering $(G^{(1)},\cdots, G^{(i)}...)$ by differential degree on a collection of ``curvature'' forms satisfying some Bianchi identity $\dd G^{(i)} = P^{(i)}(G)$. In addition to the differential degree of each of these forms
$$
g^i \, =\,  \mathrm{deg}(G^{(i)})\, \equiv \, \vert G^{(i)}\vert ,
$$
we may attempt to assign another natural number, the ``weight'', to each of these as follows. Denoting these (to be determined) numbers by
$$
w_i \equiv w(G^{(i)}) \quad \in \quad \mathbb{N}\, ,
$$
we demand that the each monomial term in the right hand side of \eqref{BianchiIdentitiesEquation} has total weight (the sum of each of the generators in the monomial) equal to that of the left hand side. That is, for
$$
\dd G^{(i)} \,=\, P^{(i)}(G)\, = \,   \sum_{\mathrm{n}\mbox{-}\mathrm{mon.}} P^{(i)}_n (G^{(1)},\cdots, G^{(i-1)}) \, ,
$$
we have 
$$
w_i = w(P^{(i)}_n)
$$
for every $n$, and then for each $i$. In particular, we also assume that the differential $\dd$ preserves weights. This demand is equivalent to a set of linear equations whose solution -- if it exists -- determines the positive weight assignment.

\begin{example}[\bf Some weight assignments]\label{ExamplesOfWeights}

\begin{itemize}
\, 

\medskip
\item We already mentioned the prime example of the $C$-field Bianchis $\dd G_4 =0,$ $\dd G_7 = \frac{1}{2} G_4 G_4$, which admit the weights $w(G_4)=3$ and $w(G_7)=6$, i.e.
$$
(w_1,w_2) \, =\, (3,6)\, .
$$
\item Next, consider example with curvatures $(F_2,F_4,F_5)$ where $\dd F_2= 0$, $\dd F_4 =0$,
$$
\dd F_5 = F_2 F_4 + (F_2)^{3}
$$
Denoting the corresponding weights by $w_1, w_2, w_3$, the homogeneity of the last equation yields the two linear equations
$$
w_3 = w_1 + w_2 \qquad , \qquad w_3 = 3 w_1 \, ,
$$
which is solved by 
$$
w_2 =2 w_1.
$$
Choosing then the lowest weight to be $w_1=1 $ yields
$$
(w_1, w_2, w_3) = (1,2,3)
$$

\item 
Consider $(F_2,H_3, F_4, K_6, K_8)$ with
$d F_2=0$, $d H_3 =0 $, $d F_4 = H_3 F_2 $,
$$
\dd K_6  \, =\, H_3 F_4 + H_3 F_2 F_2 
$$
and
$$
\dd K_8 = H_3 K_6 + H_3 (F_2)^3 + H_3 F_4 F_2 \, .  
$$
It is easy to check that this is well-defined, in that $\dd^2=0$ for arbitrary (non-trivial) solutions. 

Then, reading off the weights linear equations as above yields
\begin{align*}
w_3 &= w(F_4) = w(F_2)+ w(H_3) = w_1 + w_2 
\\ 
w_4 &= w(K_6) = w(H_3) + w(F_4) = w_2 + w_3 \\
   &\phantom{= w(K_6) \, \, } = 2w_1 + w_2 
\\
w_5 &= w(K_8) = w(H_3) + w(K_6) = w_2 + w_4 
\\ 
&\phantom{= w(K_6) \, \, } =  w(H_3) + 3\cdot w(F_2) = w_2 + 3w_1 
\\
&\phantom{= w(K_6) \, \, } = w_1 + w_2 + w_3 \, .
\end{align*}
It can be easily checked that upon choosing $w_1=w_2=1$, the full weights are given by
$$
(w_1,w_2,w_3,w_4,w_5) \, = \, (1,1,2,3,4) 
$$
\end{itemize}
\end{example}

\subsection*{Existence of positive weighted trivializations}
\label{ExistenceOfPositiveWeightAssignments}

Since our curvature forms satisfying the Bianchi identities define, and are defined by, a DGCA map 
$$
G\, : \, A=\mathrm{CE}^\bullet(\mathfrak{g}) \longrightarrow \Omega^\bullet(M)
$$ for some universally defined classifying algebra, the weight calculations above may be equivalently assigned directly on the generators of $\mathrm{CE}(\mathfrak{g})$, i.e.,  
$$
w(x^i) = w\big(\dd_{\mathrm{CE}}(x^i)\big) = w\big(P^i(x)\big) =  w\big( P^i_n(x) \big) \qquad \forall n\geq 2\, .
$$
Indeed, if such a  positive weight exists, it is a property of the target classifying DGCA.

In more detail, let
$A=\mathrm{CE}^\bullet(\frg)=(\FR[\{x^i\}_{i\in I}] ,\dd_{\mathrm{CE}})$ be a DGCA over $\FR$ of finite type, with generators $x^i$ of
positive differential degree $g^i$, and write
$\dd_{\mathrm{CE}}x^i=P^i(x)$.
A \textit{positive weight assignment} consists of integers $w_i>0$ such
that every nonzero monomial in $P^i$ has total weight $w_i$. Equivalently,
it defines a second grading of the form
\[
 A=\bigoplus_{w\geq 0}A_w,
 \qquad A_0=\mathbb{R},
 \qquad A_uA_v\subseteq A_{u+v},
 \qquad \dd_{\mathrm{CE}}A_w\subseteq A_w\, ,
\]
yielding a weight \emph{decomposition}.
Positivity here is imposed on the Chevalley-Eilenberg generators, while the usual dual convention reverses the signs of the weight on the corresponding dual $L_\infty$-algebra generators\footnote{That is, if we assume throughout that $A=\mathrm{CE}(\frg)$ is a Chevalley--Eilenberg algebra of a (connective) $L_\infty$-algebra, as we shall.} 
$$
\frg  \, \cong  \, \bigoplus_{w\leq0}^{} \frg_{w} \, . 
$$
Notice that integer positivity implies that, for any $t\in \FR$, the map $S_t: A\rightarrow A$ given by 
$$
S_t(x^i)\, :=\, t^{w_i}x^i
$$ yields DGCA endomorphisms of $A$. In particular, we have a ``scaling action'' by the multiplicative positive real line $R_{>0}$ on the weighted algebra $A$.

\begin{lemma}[\bf Criterion for positive weights]
\label{PositiveWeightCriterion}
For prescribed generators $\{x^i\}$ on $A$, a collection $w_i>0$ is a consistent weight assignment
if and only if the (degree-zero) ``Euler derivation'' on $A$
\[
 \CE\, :=\, \sum_i w_i x^i\frac{\partial}{\partial x^i} \qquad , \qquad \CE(x^i)\, = \, w_i x^i  \quad \text{(no sum)}
\]
commutes with $\dd_{\mathrm{CE}}$
$$
[\CE,\,\dd_{\mathrm{CE}}] \, = \, 0 \, .
$$
Explicitly, the condition is
\begin{equation*}
 \sum_j w_jx^j\frac{\partial P^i}{\partial x^j}\, =\, w_iP^i
 \qquad\text{for every }i.
\end{equation*}
If there are finitely many generators, this is the finite linear system
\begin{equation}\label{PositiveWeightLinearCriterion}
 w_i\,= \,\sum_j\alpha_jw_j
 \quad\text{whenever }c^i_\alpha\neq0\, ,
 \qquad w_i>0\, ,
\end{equation}
which admits positive integer solutions if and only if it admits positive
real solutions.
\end{lemma}
\begin{proof}
The commutator of the two derivations vanishes precisely when it vanishes
on each generator
\begin{align*}
[\CE,\,\dd_{\mathrm{CE}}] (x^i) &= \CE(\dd_{\mathrm{CE}}(x^i)) - \dd_{\mathrm{CE}} (\CE (x^i)) = \CE (P^i(x)) - w_i P^i \\
&=  \sum_j w_jx^j\frac{\partial P^i}{\partial x^j}-w_iP^i \, .
\end{align*}
Since $\CE(x^\alpha)=(\sum_j\alpha_jw_j)x^\alpha$, this is equivalent
to \eqref{PositiveWeightLinearCriterion}. The resulting linear solution subspace
is defined over $\mathbbm{Q}$ and so if it meets the open positive orthant, it
contains a positive rational point. Multiplying by the least common denominator gives a corresponding list of admissible positive
integer weights.
\end{proof}

More intrinsically, without choosing explicit generators, this says that $A$ admits such a weight decomposition if and only if it admits a
(locally) finite, diagonalizable degree-zero derivation $\CE$ commuting with
$\dd_{\mathrm{CE}}$ and having positive integer eigenvalues on the
indecomposables $Q(A)=A^+/(A^+)^2$.\footnote{Here \(A^+=\bigoplus_{q>0}A^q\) is the augmentation ideal, and \(Q(A)=A^+/(A^+)^2\) is the space of indecomposables, canonically identified with \(V\) for a chosen free presentation \(A=\Lambda V\).} Homogeneous lifts of an eigenbasis of
$Q(A)$ then give homogeneous free generators.\footnote{In particular,  failure of
\eqref{PositiveWeightLinearCriterion} in one presentation does not exclude non-trivial 
weight assignment after changing generators.} Requiring the derivation to preserve
the specified generator space $V$ gives the criterion for a strict weight
decomposition of the dual $L_\infty$-algebra.

\begin{remark}[\bf Sufficient conditions]\label{SufficientConditions}

Useful sufficient conditions include the following:
\begin{itemize}
\item If the differential is quadratic and every generator has degree $g^i\geq2$, then the assignment
$w_i=g^i-1$ is a consistent positive weight. Indeed, any term $\sim x^jx^k$ in $P^i(x)$ satisfies
$g^j+g^k=g^i+1$, hence the corresponding weights satisfy 
$$
w_j+w_k= g^j-1 + g^k-1 = g^i +1 -1 -1 = w^i \, .
$$
A prime example here is the rational 4-sphere minimal model  $\mathrm{CE}(\mathfrak{l} S^4) = \FR_\dd[g_4,g_7]\big/(dg_4, dg_7-\tfrac{1}{2}g_4^2)$, with weights $3$ and $6$, respectively. 
\item Positive weights extend across a cocycle extensions
$$
A \longrightarrow A[y] \big/ (dy=\omega) \, ,
$$ whenever $\omega$ is homogeneous cocycle
of positive weight $W$. Indeed, one can simply set $w(y):=W$. 
\item Any connected nilpotent rational homotopy type of finite type which
is \emph{formal} admits a positive-weight Sullivan minimal model \cite[\S~3.1]{CiriciSaleh}. Here
formality means that its minimal model is quasi-isomorphic to
$(H^*(A),0)$, not that its own differential vanishes. Again the rational $4$-sphere $\mathrm{CE}(\mathfrak{l} S^4) = \FR_\dd[g_4,g_7]\big/(dg_4, dg_7-\tfrac{1}{2}g_4^2)$ yields an example, being quasi-isomorphic to its cohomology $H^\bullet_\mathrm{CE}(\mathfrak{l} S^4) = \FR_\dd[g_4]/g_4^2$.

Another source is provided
by Sullivan models of smooth complex algebraic varieties \cite{CiriciSaleh}.
\item If $\frg$ admits a weight, then its cyclification $\mathrm{cyc}(\frg)$ and toroidification(s) $\mathrm{tor}(\frg)$ (cf. \cite{FSS17}\cite{SV1}\cite{GSS24-TDuality}) admit induced weights. More precisely, if $\mathrm{CE}^\bullet(\frg)$ admits positive
integer weights $w_i$ then it is easily checked that any
$\lambda,b$ be positive integers with $b<\lambda\min_iw_i$ yield positive integer weights on $\mathrm{CE}^\bullet(\mathrm{cyc}(\frg))$ via
\begin{align*}
 W(\omega_2)=b,\qquad W(x^i)=\lambda w_i,
 \qquad W(sx^i)=\lambda w_i-b.
\end{align*}
In particular, $$\lambda=2 \qquad \mathrm{and} \qquad b=1
$$always suffice. A straightfoward generalization holds for the toroidified algebras \cite{SV1}\cite{SV-loop}.

\end{itemize}
\end{remark}
\medskip

All rational classifying spaces determined by flux field strengths of supergravities (without cosmological constants) admit such positive-weight gradings, by appealing to their flat-model super-Minkowski cochain counterparts from which these rational spaces can be identified (or defined)\cite{FSS17}\cite{GSS24-TDuality}, in a manner analogous to that argued below for the case of 11D supergravity.

\begin{example}[\bf 4-sphere weights via super--Minkowski spacetime]\label{4SphereWeightsViaMinkowski} 
The super-Minkowski Lie algebra $\FR^{1,10\vert \bf{32}}$ is canonically weighted by assigning the ``scaling dimensions'' to its CE-generators
$$
w(e^a) = 1\qquad , \qquad w(\psi^\beta) = \frac{1}{2} \, ,
$$
which is consistent with the only (non-trivial) defining cochain differential relation 
$$
\dd e^a = (\overline{\psi} \Gamma^a \psi) \, .
$$
Then, the ``M2/M5-brane'' cochains
$$
G_4^0 \, = \, \tfrac{1}{2} (\overline{\psi}\,\Gamma_{ab} \,\psi) \, e^a e^b\qquad , \qquad G_7^0 \, = \, \tfrac{1}{5!} (\overline{\psi}\,\Gamma_{a_1\cdots a_5} \,\psi) \, e^{a_1\cdots a_5}
$$
have weights
$$
w(G_4^0) = \tfrac{1}{2} + \tfrac{1}{2} + 1 +1 = 3 \qquad \mathrm{and} \qquad w(G_7^0) = \tfrac{1}{2} + \tfrac{1}{2} + 5\cdot 1 = 6 \, ,
$$
respectively. By construction, these weights respect the differential relations satisfied by these cochains \cite{FSS17}
$$
\dd G_4^0 \, = \, 0, \qquad \dd G_7^0 \,= \,\frac{1}{2} G_4^0 \, G_4^0 \, ,
$$
and hence automatically determine the same positive weight decomposition on the target classifying algebra 
$$
\mathrm{CE}(\mathfrak{l}S^4)\, ,
$$
defined precisely by this same relation on its two generators $g_4, g_7 \in \mathrm{CE}(\mathfrak{l}S^4)$.

Essentially the same argument -- applied on sets of cochains $\{F_i=F_i(e,\psi)\}_{i\in I}$ on super-Minkowski spacetimes $\FR^{1,d\vert \bf{N}}$ that satisfy corresponding Bianchi identities (see e.g. \cite{FSS17}\cite{GSS24-TDuality}) -- determines positive weights on the corresponding classifying $L_\infty$-algebras.
\end{example}
This yields another way to identify the scale invariance (a.k.a. ``trombone symmetry'') of 11D SuGra via its (higher) super--Cartan geometric formulation from \cite[Thm. 3.1]{GSS24}.
\begin{corollary}[\bf 11D SuGra scale invariance]\label{ScaleInvarianceCorollary}
The field equations of 11D SuGra on any spacetime $M$ are invariant under the scaling action of $\FR_{>0}$ given by
$$
\mathrm{Sc}_t:(e,\,\psi,\, G_4,\, G_7) \longmapsto (t e,\, t^{\frac{1}{2}}\psi,\, t^{3} G_4,\, t^{6} G_7) \, .
$$
\end{corollary}
\begin{proof}
By \cite[Thm. 3.1]{GSS24} the on-shell field equations of 11D SuGra on $M$, are \textit{equivalent} to the condition that an extending (super-torsion free) super-spacetime $(M^s,e,\psi)$ carries super-fluxes of the form 
\begin{align*}
G_4^s \,&= \, \tfrac{1}{4!}(G_4)_{a_1\cdots a_4}e^{a_1\cdots a_4} \qquad + \qquad   \tfrac{1}{2}(\overline{\psi} \Gamma_{ab} \psi) e^{a}e^{b} \\
G_7^s \,&= \, \tfrac{1}{7!}(G_7)_{a_1\cdots a_7}e^{a_1\cdots a_7}  \qquad +\qquad   \tfrac{1}{5!}(\overline{\psi} \Gamma_{a_1\cdots a_5} \psi) e^{a_1}\cdots e^{a_5}
\end{align*}
which satisfy the same $\mathfrak{l}S^4$-Bianchi identity
$$
\dd G_4^s =0 \qquad , \qquad  \dd G_7^s \, = \, \frac{1}{2} G_4^s \wedge G_4^s \, .
$$
But under the assigned weights, these equations are by construction invariant under the corresponding scaling action, just as is the super-torsion free condition $\dd e + \omega\wedge e - (\overline{\psi}\Gamma\psi)=0$.
\end{proof}
The same superspacetime argument can be applied on to recover the scale invariance of all descendant theories from 11D, by reducing this via cyclification and toroidification(s) (see e.g. \cite{GSIIA}\cite{GSS26Survey}). Under the assumption that all other non-descendant (ungauged) supergravities have such super-flux superspace formulations, then the same argument applies for these too. 
\medskip 

Thus henceforth we shall assume the following:
\begin{assumption}[\bf Weighted Bianchi algebra]\label{ass:main}
$A=\mathrm{CE}(\frg)$ will be the finite-type Sullivan minimal model of a connected nilpotent space, equipped with a positive integer weight decomposition preserved by the differential. That is,
\[
 A=\mathrm{C E}(\frg)=\big(\FR[\{x^{(i)}\}_{i\in I}],\, \dd_{\mathrm{CE}}\big),
 \qquad |x^{(i)}|=g^i\geq1,
 \qquad \ddCE x^{(i)}=P^{(i)}(x)\, ,
\]
where the generators $\{x^{(i)}\}$ are ordered by weight, and each
$P^{(i)}$ is decomposable\footnote{Namely \(
d_{\mathrm{CE}}x^{(i)}
=P^{(i)}(x)
\) contains \textit{no linear} terms in the generators, for all $i\in I$.} and  homogeneous of strictly positive integer
weight $w_i=w(x^{(i)})$. Thus every generator occurring in $P^{(i)}$ has
strictly smaller weight than $w_i$.
\end{assumption}

\subsection*{Weighted trivializations}
\label{subsec:WeightedTrivializations}

The consistency ``integrability'' relations
$$
0= \ddCE^2 x^{(i)} = \ddCE P^{(i)}(x) = \sum_{j} \ddCE x^{(j)} \wedge \iota_{x^{(j)}} P^{(i)}(x)  \equiv \sum_{j} \ddCE x^{(j)} \wedge \partial_{j} P^{(i)}(x) \, ,
$$ 
immediately imply the following.

\begin{lemma}[\bf Universal Bianchi derivative identity]\label{lem:matrix} Denoting the \textit{Bianchi derivative matrix} by
\begin{align}\label{DerivativeMatrix}
M^{(i)}{}_{(j)} \, := \,  \iota_{x^{(j)}} P^{(i)}\, , 
\end{align}
and the differential degrees $g^i = \vert x^{(i)}\vert$ so that $g^i +1= \vert P^{(i)} \vert $, then in the setting of \ref{ass:main}, the following identity holds
\begin{equation*}
(-1)^{g^k}\ddCE M^{(i)}{}_{(k)} +  \sum_j M^{(j)}{}_{(k)}M^{(i)}{}_{(j)}
       \, =0\, .
\end{equation*}
Moreover, for the weighted Euler derivation
$\CE=\sum_j w_jx^{(j)}\partial_j$,
\begin{equation}\label{eq:euler}
 \sum_jw_jx^{(j)}M^{(i)}{}_{(j)}\, =\, \CE (P^{(i)}) \, = \,  w_iP^{(i)}.
\end{equation}
\end{lemma}
\begin{proof}
The identity $\ddCE^2x^{(i)}=0$ reads
$\sum_jP^{(j)}\partial_jP^{(i)}=0$, upon which we may apply a further derivative $\partial_k$
$$
\sum_j\partial_k(P^{(j)}\partial_jP^{(i)})\, =\, 0 \, . 
$$
Its first term is $\sum_jM^{(j)}{}_{(k)}M^{(i)}{}_{(j)}$, while
in passing through $P^{(j)}$ for the second term and  interchanging $\partial_k\partial_j$ with
$\partial_j\partial_k$ we get an overall sign
\[
 (-1)^{g^k(g^j+1)}(-1)^{g^kg^j}=(-1)^{g^k}.
\]
The second term is therefore
\[
(-1)^{g^k}\sum_jP^{(j)}\partial_jM^{(i)}{}_{(k)}\, = \, (-1)^{g^k}\ddCE M^{(i)}{}_{(k)},
\]
as claimed.
Equation \eqref{eq:euler} follows from weight homogeneity.
\end{proof}

Since any solution of the Bianchis on a spacetime $X$ via differential forms $\{G^{(i)}\}$ is identified precisely by a DGCA map
$$
G\, : \, \mathrm{CE}(\frg) \longrightarrow \Omega^\bullet(M)\, ,
$$
it follows that any such solution satisfies the induced identity with respect to the de Rham differential and wedge product
\begin{equation}\label{MClikeEquation}
(-1)^{g^k}\dd M^{(i)}{}_{(k)}(G) +  \sum_j M^{(j)}{}_{(k)}\wedge M^{(i)}{}_{(j)}
       (G)\, =\, 0\, .
\end{equation}
\medskip

\begin{proposition}[\bf Weighted trivialization of Bianchis]\label{TrivializationOfBianchis}
Let $U\hookrightarrow M$ be a (contractible) chart on $X$. Then for any collection $G=\{G^{(i)}\}$, solving \eqref{BianchiIdentitiesEquation} and ordered by weight, there exists  local potentials $\{C^{(i)} \in \Omega^{g_i-1}_\mathrm{dR}(U)\}$ which satisfy
\begin{align}\label{TrivializingGi}
G^{(i)} \, &= \, \dd C^{(i)} + R^{(i)}(C,G) \,  \nonumber 
\\ 
&:= \, \dd C^{(i)}+ \frac{1}{w_i} \sum_j w_j C^{(j)} \wedge M^{(i)}{}_{(j)}(G) 
\end{align}
\end{proposition}
\begin{proof}
This follows by induction on weight. The lowest-weight generators' Bianchi is simply their closure condition, and assume the decomposition has been achieved up to weight $w_{i-1}$. At the next weight $w_{i}$, all terms in the definition of $R^{(i)}$ involve previously treated
potentials and curvatures. Thus, computing its differential,
\begin{align*}
 \dd R^{(i)}
&=\,\frac1{w_{i}} \sum_jw_j\dd C^{(j)}M^{(i)}{}_{(j)}
  +\frac1{w_{i}}\sum_kw_k(-1)^{g^k-1}  C^{(k)}\dd M^{(i)}{}_{(k)}
 \\
 &=\,\frac1{w_{i}} \sum_jw_jG^{(j)}M^{(i)}{}_{(j)}
 -\frac1{w_i}\sum_{k,j}w_kC^{(k)}M^{(j)}{}_{(k)}M^{(i)}{}_{(j)}
 \\
 &\hspace{6mm}+\frac1{w_i}\sum_kw_k(-1)^{g^k-1}  C^{(k)}\dd M^{(i)}{}_{(k)}
\\
&=\, \frac1{w_{i}} \sum_jw_jG^{(j)}M^{(i)}{}_{(j)} \, = \,P^{(i)}(G)\, ,
\end{align*}
where the latter two terms vanish by
\eqref{MClikeEquation}, and the remaining term is precisely \eqref{eq:euler}. 

Thus $G^{(i)}-R^{(i)}$ is closed
$$
\dd G^{(i)} - \dd R^{(i)} \, = \dd G^{(i)} - P^{(i)}(G) = 0 \, , 
$$
and hence by Poincare's lemma shows that there exists a corresponding potential $C^{(i)}$. Finally note that generators with the same weight may be treated simultaneously as above.
\end{proof}

\begin{example}[\bf Examples of potentials via inductive formula]
\, 
\medskip 

\begin{itemize}
\medskip 
\item Consider the second example from Ex. \ref{ExamplesOfWeights}: $(F_2,F_4,F_5)$ where $\dd F_2= 0$, $\dd F_4 =0$,
$$
\dd F_5 = F_2 F_4 + (F_2)^{3}
$$
with weights
$$
(w_1, w_2, w_3) = (1,2,3)\, , 
$$
First we compute the Bianchi derivative matrix \eqref{DerivativeMatrix}, yielding the only non-trivial entries
\begin{align*}
M^{(3)}{}_{(1)} \, = \,F_4 + 3 F_2 F_2 \qquad , \qquad M^{(3)}{}_{(2)} \, = \,  F_2 .
\end{align*}
Then, plugging in the trivialization formula \eqref{TrivializingGi}, 
we have $F_2 = \dd A_1$, $F_4 = \dd A_3$, and
\begin{align*}
F_5 \, &= \, \dd A_4 + \frac{1}{3}\bigg( 1\cdot  A_1 (F_4 + 3 F_2 F_2 ) + 2 \cdot  A_3 F_2    \bigg) \\
&= \, \dd A_4 + \frac{1}{3} A_1 F_4 + A_1F_2 F_2 + \frac{2}{3} A_3 F_2 \, .
\end{align*}
\item Consider $(F_2,H_3, F_4, K_6, K_8)$ with
$d F_2=0$, $d H_3 =0 $, $d F_4 = H_3 F_2 $,
$$
\dd K_6  \, =\, H_3 F_4 + H_3 F_2 F_2 
$$
and
$$
\dd K_8 = H_3 K_6 + H_3 (F_2)^3 + H_3 F_4 F_2 \, .  
$$ 
with 
$$
(w_1,w_2,w_3,w_4,w_5) \, = \, (1,1,2,3,4) \, .
$$
The Bianchi derivative matrix has the following non trivial entries:
\begin{align*}
    M^{(3)}{}_{(1)} \, &= \, H_3  \qquad , \qquad M^{(3)}{}_{(2)} \, = \,  F_2 \\
M^{(4)}{}_{(1)} \, &= \, 2 H_3 F_2, \qquad , \qquad M^{(4)}{}_{(2)} \, = \,  F_4 + F_2 F_2, \qquad , \qquad M^{(4)}{}_{(3)} \, = \,  H_3
\\
M^{(5)}{}_{(1)} \, &= \, 3 H_3 F_2 F_2 + H_3 F_4 , \qquad , \qquad M^{(5)}{}_{(2)} \, = \, K_6 + (F_2)^3 + F_4 F_2, \\  M^{(5)}{}_{(3)} \, &= \,  H_3 F_2  , \qquad , \qquad M^{(5)}{}_{(4)} \, = \,  H_3 \, .
\end{align*}
With these, the inductive formula \eqref{TrivializingGi} gives the corresponding potentials as:
\begin{align*}
 F_2& \, = \, \dd A_1 \, , \quad \quad  H_3 \, = \, \dd B_2,\nonumber\\
 F_4& \, = \, \dd A_3+\frac12\big(A_1H_3+B_2F_2\big) \, ,\\
 K_6& \, = \, \dd A_5+\frac13\big(2A_1H_3F_2
                    +B_2(F_4+F_2^2)+2A_3H_3\big) \, ,\\
 K_8& \, = \, \dd A_7+\frac14\big(A_1(3H_3F_2^2+H_3F_4)
                    +B_2(K_6+F_2^3+F_4F_2)\nonumber\\
 &\hspace{36mm}+2A_3H_3F_2+3A_5H_3\big) \, .
\end{align*}
As a direct check
of the first nonclosed case,
\[
 \dd\left(\frac12(A_1H_3+B_2F_2)\right) \, = \, H_3F_2 \, .
\]
For the next case, the terms involving $B_2H_3F_2$ from
$\dd(B_2F_4)$ and $2\dd(A_3H_3)$ cancel, leaving
$\dd {K_6}=H_3F_4+H_3F_2^2$.
\end{itemize}
\end{example}

\subsection*{Gauge transformations between weighted trivializations}
\label{subsec:WeightedHigherGauge}

We retain the hypotheses and notation of
Prop.~\ref{TrivializationOfBianchis}, working on a contractible chart
$U$. For $G$ a fixed solution of the Bianchis we use the same expression
\[
 R^{(i)}(a,G):=\frac{1}{w_i}\sum_jw_j a^{(j)}\wedge M^{(i)}{}_{(j)}(G)
\]
for tuples $a$ of shifted lower differential degrees, with the displayed
factor order.

\begin{corollary}[\bf Local gauge transformations at fixed curvature]
\label{cor:WeightedGaugeTrivializations}
Let $C$ and $\hat C$ be two weighted trivializations of the same
curvature $G$
\begin{align*}
 G^{(i)}\,&=\, \dd C^{(i)}+R^{(i)}(C,G) 
 \\ 
        &=\,  \dd\hat C^{(i)}+R^{(i)}(\hat C,G)\, .
\end{align*}
Then:
\begin{itemize}
\item[\bf (i)]
For curvatures of differential degrees with $\vert G^{(i)}\vert=g^i\geq2$ for every $i$, there exist forms 
$\Lambda^{(i)}\in\Omega^{g^i-2}_{\mathrm{dR}}(U)$  constructed by induction on weight such that
\begin{align}\label{eq:WeightedFiniteGauge}
 C^{(i)}-\hat C^{(i)}
 \,&=\, \dd\Lambda^{(i)}
  -\frac{1}{w_i}\sum_jw_j\Lambda^{(j)}\wedge M^{(i)}{}_{(j)}(G) \\
  &=\, \dd \Lambda^{(i)} -R^{(i)}(\Lambda,G) \, . \nonumber
\end{align}
Allowing for curvature differential degree $g^i=1$, the corresponding (gauge) parameters exist if and only if
$C^{(i)}=\hat C^{(i)}$ for every degree-one generator, with negative-degree
parameter forms are understood to vanish.
\item[\bf (ii)]
Changing a trivialization $C$ by the right-hand side of
\eqref{eq:WeightedFiniteGauge}, for any such tuple $\Lambda$, preserves
its curvature $G$ via \eqref{TrivializingGi}.
\end{itemize}
\end{corollary}

\begin{proof}
Denote the difference of trivializations by $a^{(i)}:=C^{(i)}-\hat C^{(i)}$, so that subtracting the two
trivialization formulas gives
\[
 \dd a^{(i)}\, =\, -\frac{1}{w_i}\sum_jw_j a^{(j)}\wedge M^{(i)}{}_{(j)}\, .
\]
Suppose the gauge parameters $\Lambda$ have been constructed at all lower weights up to $w^{i-1}$, and hence let
\[
 Q^{(i)}\, :=\, a^{(i)}
       +\frac{1}{w_i}\sum_jw_j\Lambda^{(j)}\wedge M^{(i)}{}_{(j)}\, .
\]
Calculating the differential and substituting the lower-weight equations yields
\begin{align*}
 \dd Q^{(i)}
 &=\frac{1}{w_i}\sum_jw_j
      \big(\dd\Lambda^{(j)}-a^{(j)}\big)\wedge M^{(i)}{}_{(j)}
   +\frac{1}{w_i}\sum_kw_k(-1)^{g^k-2}
      \Lambda^{(k)}\wedge\dd M^{(i)}{}_{(k)}\\
 &=\frac{1}{w_i}\sum_kw_k\Lambda^{(k)}\wedge
   \left(
     \sum_jM^{(j)}{}_{(k)}\wedge M^{(i)}{}_{(j)}
     +(-1)^{g^k}\dd M^{(i)}{}_{(k)}
   \right)=0.
\end{align*}
The last equality is the universal Bianchi derivative identity from Lem. \ref{lem:matrix} evaluated
on $G$ \eqref{MClikeEquation}. If $g^i\geq2$, the form $Q^{(i)}$ has positive degree, so
the Poincar\'e lemma gives $Q^{(i)}=\dd\Lambda^{(i)}$.

If the curvature differential degree is one, $g^i=1$, then
$M^{(i)}{}_{(j)}(G)$ can be nonzero only for $g^j=1$. Consequently
$R^{(i)}(\Lambda,G)=0$, since the relevant gauge parameters have degree
$-1$. Equation~\eqref{eq:WeightedFiniteGauge} therefore requires
$a^{(i)}=0$.

Finally, applying $\dd+R(-,G)$ to the
right-hand side of \eqref{eq:WeightedFiniteGauge} gives zero by the
same matrix identity, proving the converse.
\end{proof}
Fix a curvature $G$ and write $\Lambda_0 \, := \, C$ for its trivializing gauge potentials. Let the ``level-$r$ gauge parameters'' have components
\[
 \Lambda_r^{(i)}\in\Omega^{g^i-1-r}_{\mathrm{dR}}(U) \, ,
 \qquad r\geq0 \, ,
\]
and define the operator on tuples of level $r$
\begin{equation}\label{eq:WeightedGaugeDifferential}
 \DD^G_r \, := \, \dd+(-1)^rR(-,G) \, ,
 \qquad
 (\DD^G_r\Lambda_r)^{(i)}
 \, = \, \dd\Lambda_r^{(i)}
   +\frac{(-1)^r}{w_i}
        \sum_jw_j\Lambda_r^{(j)}\wedge M^{(i)}{}_{(j)}(G) \, .
\end{equation}
Thus $\DD^G_r$ raises differential degree by one and maps level $r$
to level $r-1$ and at level zero 
$$\DD^G_0C \, =  G \, .$$
The signs in the transformation formulas therefore alternate:
\begin{align*}
 \DD^G_1\Lambda_1
 \, &= \, \dd\Lambda_1-R(\Lambda_1,G) \, ,\\
\DD^G_2\Lambda_2
 \, &= \, \dd\Lambda_2+R(\Lambda_2,G) \, ,\\
 \DD^G_3\Lambda_3
 \, &= \, \dd\Lambda_3-R(\Lambda_3,G) \, ,
\end{align*}
and so on.
\begin{corollary}[\bf Higher gauge transformations at fixed curvature]
\label{cor:WeightedHigherGauge}
\,
\begin{itemize}

\item[\bf (i)] The operators from \eqref{eq:WeightedGaugeDifferential} satisfy
\begin{equation}\label{eq:WeightedGaugeComplex}
 \DD^G_r\DD^G_{r+1} \, = \, 0\qquad(r\geq0) \, .
\end{equation}

\item[\bf (ii)]
The successive finite gauge and gauge-for-gauge transformations
between parallel level-$r$ data are parametrized by level $r+1$:
\begin{equation}\label{eq:WeightedHigherFiniteGauge}
 \Lambda_r^{(i)}-\hat\Lambda_r^{(i)}
 \, = \, (\DD^G_{r+1}\Lambda_{r+1})^{(i)}
 \, = \, \dd\Lambda_{r+1}^{(i)}
  +\frac{(-1)^{r+1}}{w_i}
       \sum_jw_j\Lambda_{r+1}^{(j)}\wedge M^{(i)}{}_{(j)}(G) \, .
\end{equation}
Here parallel means that the lower source and target data are fixed, so that
$\DD^G_r(\Lambda_r-\hat\Lambda_r)  = 0 .$
In particular, for $r  =  0$, this is precisely the condition that $C$ and
$\hat C$ trivialize the same curvature $G$, while for $r=1$, this is the condition that $\Lambda_1$ and $\hat{\Lambda}_{1}$ are gauge transformations between the same two potentials $C$ and 
$\hat{C}$, and so on.
\item[\bf (iii)] 
For any such \(\DD^G_r\)-closed difference $\DD^G_r(\Lambda_r-\hat\Lambda_r)  = 0$, the higher gauge parameters in
\eqref{eq:WeightedHigherFiniteGauge} always exist (locally) if all component differential degrees $\vert G^{(i)}\vert = g^i$ satisfy 
$$ 
g^i\geq r+2 \, . 
$$  
Otherwise, these exist if and only if
\begin{equation}\label{eq:WeightedHigherDegreeZero}
 \Lambda_r^{(i)} \, = \, \hat\Lambda_r^{(i)}
 \qquad\text{for every }i\text{ with }g^i \, = \, r+1 \, .
\end{equation}

\end{itemize}
\end{corollary}

\begin{proof}
Applying \eqref{eq:WeightedGaugeDifferential} twice, the terms
containing $\dd\Lambda_{r+1}^{(j)}$ cancel, and the remaining terms are
\[
 (\DD^G_r\DD^G_{r+1}\Lambda_{r+1})^{(i)}
 \, = \, -\frac{1}{w_i}\sum_kw_k\Lambda_{r+1}^{(k)}\wedge
   \left(
     \sum_jM^{(j)}{}_{(k)}\wedge M^{(i)}{}_{(j)}
     +(-1)^{g^k}\dd M^{(i)}{}_{(k)}
   \right) \, = \, 0 \, .
\]
This proves \eqref{eq:WeightedGaugeComplex}, so every transformation
\eqref{eq:WeightedHigherFiniteGauge} preserves the prescribed lower
boundary.

Conversely, write $a \, := \, \Lambda_r-\hat\Lambda_r \, $ with
$\DD^G_r a \, = \, 0 \, ,$ and assume the construction is complete at
lower weights. Consider the form
\[
 Q_{r+1}^{(i)} \, := \, a^{(i)}+(-1)^rR^{(i)}(\Lambda_{r+1},G)
\]
of degree $g^i-1-r \, .$ The calculation in the preceding proof, using
$|\Lambda_{r+1}^{(k)}| \, = \, g^k-2-r \, ,$ gives
\[
 \dd Q_{r+1}^{(i)}
 \, = \, \frac{1}{w_i}\sum_kw_k\Lambda_{r+1}^{(k)}\wedge
   \left(
     \sum_jM^{(j)}{}_{(k)}\wedge M^{(i)}{}_{(j)}
     +(-1)^{g^k}\dd M^{(i)}{}_{(k)}
   \right) \, = \, 0 \, .
\]

When $g^i \, = \, r+1 \, ,$ decomposability gives
$M^{(i)}{}_{(j)} \, = \, 0$ for $g^j\geq r+2 \, ,$ while
$\Lambda_{r+1}^{(j)} \, = \, 0$ for $g^j\leq r+1 \, .$ Hence
$R^{(i)}(\Lambda_{r+1},G) \, = \, 0$ and the equation is precisely
$a^{(i)} \, = \, 0 \, ,$ explaining both necessity and sufficiency of
\eqref{eq:WeightedHigherDegreeZero}. Components with $g^i<r+1$
vanish for degree reasons. This completes the induction.
\end{proof}

Note that the operators $\DD_r^G$ from \eqref{eq:WeightedGaugeDifferential} are defined globally, and still satisfy \eqref{eq:WeightedGaugeComplex}. This implies that differences of (local or global) weighted trivializations and higher gauge parameters of a fixed curvature $G$ fit into a cochain \textit{gauge} complex (even if some of these do not exist).
\begin{definition}[\bf Weighted gauge complex at fixed curvature]
\label{WGPC:def:gaugecomplex}
Let
\[
 \mathcal A_r(M)\, := \,\prod_i\Omega^{g^i-1-r}(M)\, ,
\]
with negative-degree components understood to vanish. 
The \emph{weighted gauge complex} of $G$ on $M$ is the nonpositively graded
cochain complex $(L^\bullet,\partial_L)$ given by
\[
 L^0\, := \,\ker\DD_0^G\, ,\qquad
 L^{-r}\, := \,\mathcal A_r(M)\quad(r\geq1)\, ,\qquad
 \partial_L|_{L^{-r}}\, := \,\DD_r^G\quad(r\geq1) \, ,
\]
with $L^j=0$ for $j>0$ and zero differential on $L^0$. Thus
\[
 \cdots\xrightarrow{\DD_3^G}\mathcal A_2(M)
 \xrightarrow{\DD_2^G}\mathcal A_1(M)
 \xrightarrow{\DD_1^G}\ker\DD_0^G\longrightarrow0 \, .
\]
\end{definition}
This is a complex by $\DD_r^G\DD_{r+1}^G=0$. Its degree-zero elements
are homogeneous differences of potentials, i.e., if a trivialization $C_*$
exists, every such element $c$ is the difference $(C_*+c)-C_*$ of
two trivializations of $G$. Similarly, degree $-r$ consists of level-$r$ gauge parameters.

\paragraph{The first two morphism levels.}
The trivializing potentials $C$ and $\hat C$ may be thought of as two parallel
null-transformations $0\to G$. Two gauge-transformations with the same boundary potentials
are then parallel $2$-morphisms, as depicted below:
\begin{center}
\begin{tikzpicture}[
  >=Stealth,
  curvature path/.style={->,line width=.55pt},
  transformation/.style={double,double distance=1.2pt,
    line width=.4pt,-{Implies[length=5pt,width=5pt]}}
]
  \node (zero) at (-2.6,0) {$0$};
  \node (flux) at (2.6,0) {$G$
  \, .};
  \draw[curvature path]
    (zero) .. controls (-1.3,1.5) and (1.3,1.5) ..
    node[pos=.5,above=4pt] {$C$} (flux);
  \draw[curvature path]
    (zero) .. controls (-1.3,-1.5) and (1.3,-1.5) ..
    node[pos=.5,below=4pt] {$\hat C$} (flux);
  \draw[transformation]
    (-.18,-.88) .. controls (-.85,-.4) and (-.85,.4) ..
    node[pos=.5,left=5pt] {$\Lambda_1$} (-.18,.88);
  \draw[transformation]
    (.18,-.88) .. controls (.85,-.4) and (.85,.4) ..
    node[pos=.5,right=5pt] {$\hat\Lambda_1$} (.18,.88);
\end{tikzpicture}
\end{center}
The upward orientation of the $2$-morphisms corresponds to
$\Lambda_1,\hat\Lambda_1:\hat C\Rightarrow C$, consistently with
\[
 C-\hat C \, = \, \DD^G_1\Lambda_1
 \, = \, \DD^G_1\hat\Lambda_1 \, .
\]

\paragraph{Linear composition of finite transformations.}

At fixed $G$, let $C_0,C_1,C_2$ be three weighted trivializations and
let $\Lambda_1:C_0\to C_1$ and $\Xi_1:C_1\to C_2$ be transformation
parameters, meaning
\[
 C_1-C_0 \, = \, \DD^G_1\Lambda_1 \, ,\qquad
 C_2-C_1 \, = \, \DD^G_1\Xi_1 \, .
\]
By linearity of the operator $\DD_1^G$,
\[
 C_2-C_0 \, = \, \DD^G_1(\Lambda_1+\Xi_1) \, .
\]
Their composite in the algebraic parameter description is therefore
\begin{equation*}
 \Xi_1\circ\Lambda_1 \, := \, \Lambda_1+\Xi_1\quad :\quad C_0\longrightarrow C_2
 \, .
\end{equation*}
The following triangle depicts this composition:
\begin{center}
\begin{tikzpicture}[>=Stealth, every node/.style={inner sep=3pt}]
 \node (c0) at (-2,0) {$C_0$};
 \node (c1) at (2,0) {$C_1$};
 \node (c2) at (0,-1.6) {$C_2$};
 \draw[->] (c0) -- node[above] {$\Lambda_1$} (c1);
 \draw[->] (c1) -- node[below right] {$\Xi_1$} (c2);
 \draw[->] (c0) -- node[below left] {$\Lambda_1+\Xi_1$} (c2);
\end{tikzpicture}
\end{center}
These arrows are $1$-morphisms in the trivialization groupoid, hence
$2$-morphisms between the corresponding arrows $0\to G$ in the
ambient curvature groupoid. The identity parameter at $C$ is zero,
and the inverse parameter is the negative:
\[
 \mathrm{id}_C \, := \, 0\quad : \quad C\longrightarrow C \, ,\]
 \[
 (\Lambda_1:C_0\to C_1)^{-1}
 \, := \, -\Lambda_1\quad :\quad C_1\longrightarrow C_0 \, .
\]
Addition makes this parameter composition strictly associative and
unital. 

More generally, for any level $r$, if parallel data
$A_r,B_r,C_r$ and parameters $L_{r+1},M_{r+1}$ satisfy
\[
 B_r-A_r \, = \, \DD^G_{r+1}L_{r+1} \, ,\qquad
 C_r-B_r \, = \, \DD^G_{r+1}M_{r+1} \, ,
\]
then their successive composition at this level is
\[
 M_{r+1}\circ L_{r+1}
 \, := \, L_{r+1}+M_{r+1}\quad :\quad A_r\longrightarrow C_r \, ,
\]
since $  C_r-A_r  =  \DD^G_{r+1}(L_{r+1}+M_{r+1})$. Identities and inverses are again zero and negation. This formula thus
specifies composition along the common level-$r$ boundary.

Indeed, these can be assembled into a definition of an actual $\infty$-groupoid of higher trivializing gauge parameters, encoding these compositions and their iterations -- up to further higher gauge transformations.
\begin{definition}[\bf Infinity groupoid of weighted trivializations]
\label{WGPC:def:parameters}\label{TrivializationsGroupoid}
The \textit{simplicial set of trivializations} of a curvature $G$ on a manifold M
$$
\mathrm{Triv}\,\equiv \, \mathrm{Triv}_{W}^\infty(G)(M)
$$
has degree-$n$ set
\begin{equation}\label{WGPC:eq:compatibility}
 (\mathrm{Triv})_n\, := \,
 \left\{
 (\Lambda_r^{a_0\cdots a_r})_{\substack{0\leq r\leq n\\0\leq a_0<\cdots<a_r\leq n}}
 \;\middle|\;
 \begin{gathered}
 \Lambda_r^{a_0\cdots a_r}\in\mathcal A_r(M)\, ,
 \qquad \DD_0^G\Lambda_0^a\, = \,G\, ,\\
 \displaystyle
 (\delta\Lambda_{r-1})^{a_0\cdots a_r}
\, := \, \sum_{j=0}^r(-1)^j
 \Lambda_{r-1}^{a_0\cdots\widehat{a_j}\cdots a_r}
 \, = \,(-1)^{r+1}\DD_r^G\Lambda_r^{a_0\cdots a_r}
 \quad(1\leq r\leq n)
 \end{gathered}
 \right\} \, .
\end{equation}
Here $\Lambda_0^a=C_a$ is the vertex potential, and the conditions
hold on every ordered face of the indicated dimension.
For a monotone map $\theta:[m]\to[n]$,
pullback is defined by
\[
 (\theta^*C)_a\, = \,C_{\theta(a)}\, ,\qquad
 (\theta^*\Lambda)_r^{a_0\cdots a_r}\, = \,
 \begin{cases}
 \Lambda_r^{\theta(a_0)\cdots\theta(a_r)},
       &\theta(a_0)<\cdots<\theta(a_r),\\
 0,&\text{otherwise}.
 \end{cases}
\]
\end{definition}
Thus face maps restrict the assigned data, while degeneracies duplicate
vertices and assign zero to collapsed positive-dimensional faces.
Note that an $n$-simplex includes its entire compatible boundary data, not only
a single parameter $\Lambda_n$. Here the first three compatibility equations are
\[
 \begin{aligned}
 C_b-C_a&\, = \,\DD_1^G\Lambda_1^{ab}\, ,\\
 \Lambda_1^{ac}-\Lambda_1^{ab}-\Lambda_1^{bc}
   &\, = \,\DD_2^G\Lambda_2^{abc}\, ,\\
 \Lambda_2^{bcd}-\Lambda_2^{acd}+\Lambda_2^{abd}-\Lambda_2^{abc}
   &\, = \,\DD_3^G\Lambda_3^{abcd} \, ,
 \end{aligned}
\]
which describe arrows, transformations from an additive composite to
a direct arrow, and the next coherence, respectively. Explicitly, for instance, the middle equation here says that $\Lambda_1^{ac}$ is \textit{almost} the strict composite of $\Lambda^{ab}_1$ and $\Lambda^{bc}_1$ -- up to a higher gauge transformation $\DD^G_2 \Lambda_2^{abc}$
\begin{center}
\begin{tikzpicture}[scale=1,>=Stealth]
\node (a) at (-2,0) {$C_a$};
\node (b) at (0,2) {$C_b$};
\node (c) at (2,0) {$C_c$\, ,};
\draw[->] (a) -- node[above left] {$\Lambda_1^{ab}$} (b);
\draw[->] (b) -- node[above right] {$\Lambda_1^{bc}$} (c);
\draw[->] (a) -- node[below] {$\Lambda_1^{ac}$} (c);
\draw[double,double distance=2pt,-{Implies[length=5pt,width=7pt]}]
 (0,1.25) -- node[right=5pt] {$\Lambda_2^{abc}$} (0,0.3);
\end{tikzpicture}
\end{center}
and so on, as intuitively expected by an $\infty$-groupoid encoding (higher) gauge transformations. 

If a vertex
$C_*$ exists, subtraction from every vertex turns $\Triv$
into a simplicial vector space and hence it is Kan (cf. Lem. \ref{prop:WeightedTrivKan} below). If no vertex exists,
it is empty.

\begin{remark}[\bf Relation to the Dold--Kan construction]
\label{WGPC:rem:doldkan}
Regard the gauge complex (Def. \ref{WGPC:def:gaugecomplex}) as a chain complex in nonnegative degrees,
with $\mathcal A_r(M)$ in degree $r\geq1$, $\ker\DD_0^G$ in degree
zero, and differential $(-1)^{r+1}\DD_r^G$ from degree $r$.
After choosing a trivialization $C_*$, its Dold--Kan construction $\Gamma$ gives 
\begin{align*}(\mathrm{Triv})_n\quad&\cong_{C_*}\quad \Gamma\big(\cdots\xrightarrow{\DD_3^G}\mathcal A_2(M)
 \xrightarrow{-\DD_2^G}\mathcal A_1(M)
 \xrightarrow{\DD_1^G}\ker\DD_0^G
 \big)_n
 \\
 &:= \,\,  \quad  \operatorname{Hom}_{\mathrm{Ch}_\FR}\!\left(
 N_*(\Delta[n]),\,
 \cdots\xrightarrow{\DD_3^G}\mathcal A_2(M)
 \xrightarrow{-\DD_2^G}\mathcal A_1(M)
 \xrightarrow{\DD_1^G}\ker\DD_0^G
 \right) \, ,
\end{align*}
where $N_*(\Delta[n])$ denotes normalized simplicial chains,
with the alternating boundary. This is the classical
Dold--Kan construction \cite[\S6,8]{Kan1958}, where the isomorphism assigns to each simplex
the chain map
\[
 [a]\longmapsto C_a-C_*\, ,\qquad
 [a_0\cdots a_r]\longmapsto\Lambda_r^{a_0\cdots a_r}\quad(r\geq1) \, ,
\]
because normalized simplex chains are freely generated by the strictly increasing faces \([a_0\cdots a_r]\).

Indeed, the chain-map equations are exactly
\eqref{WGPC:eq:compatibility}, and precomposition gives the stated
faces and degeneracies. Multiplication in degree $r$ by
$(-1)^{r(r-1)/2}$ identifies the displayed chain complex with the
one having differential $\DD_r^G$. Without choosing $C_*$, a nonempty
$\Triv$ is canonically a torsor for the displayed Dold--Kan
simplicial vector space: it acts freely and transitively in each
degree by addition of homogeneous parameter data.
\end{remark}

\section{An equivalent total-form model}
\label{WGPC:subsection}
We now give
a different simplicial set of trivializations and their higher gauge
transformations
in terms of differential forms on parameter simplices, which serves as a useful intermediate model towards the comparison to null-concordances. This is straightforward to write down, and absorbs the explicit combinatorial nature of Def. \ref{WGPC:def:parameters}. Nevertheless, in general, this turns out to be somewhat larger: it is \textit{not strictly} isomorphic, but rather (homotopy) equivalent to the former (Thm. \ref{WGPC:prop:comparison} below). 

\begin{definition}[\bf Infinity groupoid of weighted total-form trivializations]
\label{WGPC:def:forms}\label{TotalTrivializations}
The \textit{simplicial set of total-form trivializations} of a curvature $G$ on a manifold $M$ 
$$
\Triv_{\mathrm{total}}\, \equiv \, \mathrm{Triv}_{W,\mathrm{total}}^\infty(G)(M)
$$
has degree-$n$ set
\[
 (\Triv_{\mathrm{total}})_n\, := \,
 \left\{\mathcal C\in\prod_i\Omega^{g^i-1}(M\times\Delta^n)
 \;\middle|\;
 \dd_{M\times\Delta^n}\mathcal C
       +R(\mathcal C,\operatorname{pr}_M^*G)
       =\operatorname{pr}_M^*G\right\} \, ,
\]
where we use smooth differential forms
up to the boundary of the simplex.
\end{definition}
Note that writing $\mathcal C=\sum_{r=0}^n\mathcal C_{[r]}$ for the decomposition by simplex degree of any such total-form on $M\times \Delta^n$, so that 
the $M$-coefficient of $\mathcal C_{[r]}^{(i)}$ has degree $g^i-1-r$ on $M$,
the total equation from Def. \ref{WGPC:def:forms} is then equivalent to
\begin{equation}\label{TotalFormsEquivalentEquation}
 \DD_0^G\mathcal C_{[0]}\, = \,G\, ,\qquad
 \dd_\Delta\mathcal C_{[r-1]}+(-1)^r\DD_r^G\mathcal C_{[r]}
 \, = \,0\qquad(1\leq r\leq n) \, .
\end{equation}
For a monotone map $\theta:[m]\to[n]$, let
$\theta_*:\Delta^m\to\Delta^n$ be the associated affine map, with the corresponding
simplicial operator 
\[
 \theta^*\mathcal C
 \, := \, (\mathrm{id}_M\times\theta_*)^*\mathcal C \, .
\]
All restrictions are pullbacks of total forms, including their
components in the simplex directions. We shall suppress the pullback $\operatorname{pr}_M^*$ along the projection to $M$ in formulas below. The fact that this simplicial set is Kan follows by its (affine) linearity.

\begin{lemma}[\bf Kan property by affine linearity]
\label{prop:WeightedTrivKan}
The simplicial sets $\mathrm{Triv}_{W}^\infty(G)(M)$ and $\mathrm{Triv}^\infty_{W,\mathrm{total}}(G)(M)$ are Kan.
\end{lemma}
\begin{proof} We consider this for the latter case only, the former being analogous again by its affine linearity.
Suppose first that a trivialization of $G$ exists and choose one,
denoted $C_*$, so that it satisfies
\[
 \dd C_*^{(i)}+R^{(i)}(C_*,G) \, = \, G^{(i)} \, .
\]
For each $n$, let $\operatorname{pr}_M:M\times\Delta^n\longrightarrow M$ be the
projection, and put
\[
 C_{*,n} \, := \, \operatorname{pr}_M^*C_* \qquad\text{and}\qquad
 G_n \, := \, \operatorname{pr}_M^*G \, .
\]
Thus $C_{*,n}$ is constant in the simplex directions and has no
simplex differential legs, hence a particular $n$-simplex of trivializations.

For an arbitrary $n$-simplex $\mathcal C$, subtraction means ordinary
componentwise subtraction of differential forms on the same product,
\[
 a^{(i)} \, := \, \mathcal C^{(i)}-C_{*,n}^{(i)} \, .
\]
Both tuples $\mathcal{C}$ and $C_*$ satisfy the same inhomogeneous equation with right-hand
side $G_n$, and since $R(-,G_n)$ is linear in its first argument,
subtracting these equations gives
\[
 \dd a^{(i)}+R^{(i)}(a,G_n) \, = \, 0 \, \, .
\]
and conversely, adding $C_{*,n}$ to any solution of this homogeneous
equation gives a trivialization of $G_n$. We therefore have mutually
inverse bijections
\[
 \mathrm{Triv}(G)^\infty_{W,n}\longleftrightarrow Z_n,
 \qquad \mathcal C\longmapsto\mathcal C-C_{*,n},
 \qquad a\longmapsto C_{*,n}+a \, ,
\]
where
\[
 Z_n \, := \, \left\{a\in\prod_i\Omega^{g^i-1}(M\times\Delta^n)
 \;\middle|\; \dd a+R(a,G_n)=0\right\} \, .
\]
This solution set is a vector space, since its equation is homogeneous
and linear. In other words, choosing $C_*$ determines an origin in the
affine space of trivializations.

Faces and degeneracies are pullbacks, hence linear, preserve the
homogeneous equation, and carry $C_{*,n}$ to the corresponding reference
simplex. In particular, subtraction commutes with every face and
degeneracy. The displayed bijections therefore identify the entire
simplicial set with the underlying simplicial set of the simplicial
vector space $Z_\bullet$.

Thus, we can appeal to the standard constructive proof that every simplicial
abelian group, and hence every simplicial vector space, is Kan \cite[Lem.~7.6]{Kan1958}(see also \cite[Prop.~1.2.5.9]{Kerodon}). If a trivialization does not exist, the statement is vacuously true.
\end{proof}

\medskip
{\bf Comparing the two models.}
\medskip

The two
models have the same vertices, but their higher simplices are
different data. A total-form simplex has arbitrary smooth dependence
on simplex coordinates while a gauge-parameter simplex just records one
parameter on each face. We first describe integration and normalized
reconstruction in degrees $1$ and $2$ as operations between the two, before giving the general
comparison proposition.

\paragraph{The comparison in degree $1$.}
Writing a total-form simplex as
\[
 \mathcal C\, = \,C(t)+\dd t\,\lambda_1(t) \, ,
\]
its defining equation is equivalent to
\[
 \DD_0^G C(t)\, = \,G\, ,\qquad
 \partial_t C(t)\, = \,\DD_1^G\lambda_1(t) \, .
\]
Define
\begin{align}\label{Degree1Integration}
 I_1(\mathcal C)\, := \,
 \left(C_0=C(0),\ C_1=C(1),\
 \Lambda_1^{01}=\int_0^1\lambda_1(t)\,\dd t\right) \, .
\end{align}
Integration of the flow equation of $C(t)$ thus gives 
$$
C_1-C_0=\DD_1^G\Lambda_1^{01}\, .
$$
so that $I_1(\mathcal{C})$ is a 1-simplex of $\Triv$.

Conversely, define from any $1$-simplex of $\Triv$ the ``normalized representative''
\begin{align}\label{Degree1Whitney}
W_1(C_0,C_1,\Lambda_1^{01})
 \, := \, (1-t)C_0+tC_1+\dd t\,\Lambda_1^{01} \, .
\end{align}
It follows easily that this is a total-form simplex and moreover $$
I_1W_1=\mathrm{id}\,.
$$
Generally, however, the converse composition is not identity
$$
W_1 I_1 \, \neq \, \mathrm{id}\, .
$$
Indeed, for a
$1$-simplex
\[
 \mathcal C \, = \, C(t)+\dd t\,\lambda(t),
 \qquad
 \partial_tC(t) \, = \, \DD_1^G\lambda(t)
 \, ,
\]
with $C_0=C(0)$ and $C_1=C(1)$, its ``Whitney reconstruction'' is
\[
 W_1I_1(\mathcal C)
 \, = \,
 (1-t)C_0+tC_1+\dd t\,\overline\lambda,
 \qquad
 \overline\lambda \, := \, \int_0^1\lambda(u)\,\dd u
 \, .
\]
Thus the reconstruction replaces $C(t)$ by its affine
interpolation and $\lambda(t)$ by its constant average.

The two simplices are nevertheless \textit{homotopic} relative
to their endpoints. To see this, define
\[
 b(t)
 \, := \,
 \int_0^t\lambda(u)\,\dd u-t\overline\lambda
 \, ,
\]
(cf. \cite[Section 1.1, p. 61]{BandieraSchaetz}.) Integrating the interval equations for $\mathcal{C}$ gives
\[
 \DD_1^G b(t)
 \, = \,
 C(t)-\bigl((1-t)C_0+tC_1\bigr),
 \qquad
 \partial_t b(t)
 \, = \,
 \lambda(t)-\overline\lambda
 \, .
\]
Consequently, these combine to
\begin{align}\label{1SimplexTotalFailure}
 \dd_{M\times I_t}b-R(b,G)
 \, = \,
 \mathcal C-W_1I_1(\mathcal C)
 \, .
\end{align}
The scalar interpolation between the simplices $\mathcal{C}$ and $W_1 I_1 \mathcal{C}$
$$
(1-s)C(t)+s\bigl((1-t)C_0+tC_1\bigr)+\dd t\bigl((1-s)\lambda(t)+s\overline\lambda\bigr)
$$
therefore becomes a trivialization on the product
with $I_s$ after adding the correction $-\dd s\,b$:
\[
 \begin{aligned}
 \mathcal H
 \, := \,&
 (1-s)C(t)+s\bigl((1-t)C_0+tC_1\bigr)\\
 &+\dd t\bigl((1-s)\lambda(t)+s\overline\lambda\bigr)
 -\dd s\,b(t)
 \, .
 \end{aligned}
\]
Indeed,
\[
 \dd_{M\times I_t\times I_s}\mathcal H+R(\mathcal H,G)
 \, = \, G
 \, ,
\]
since the correction cancels the additional
$\dd s$ term produced by the interpolation.
The restrictions at $s=0,1$ are respectively
$\mathcal C$ and $W_1I_1(\mathcal C)$, while
$b(0)=b(1)=0$ ensures that the endpoint
trivializations $C_0,C_1$ remain fixed.

\paragraph{The comparison in degree $2$.}
On the geometric 2-simplex $\Delta^2\hookrightarrow \FR^3$, use coordinates $u=u_1$, $v=u_2$, $u_0=1-u-v$, oriented by
$\dd u\wedge\dd v$ and expand
\[
 \mathcal C\, = \,C(u,v)+\dd u\,\lambda_u+\dd v\,\lambda_v
                       +\dd u\wedge\dd v\,\mu_2 \, .
\]
The equations from the right hand side of Def. \ref{WGPC:def:forms} then expand as
\[
 \begin{aligned}
 \DD_0^G C&\, = \,G\, ,\\
 \partial_uC&\, = \,\DD_1^G\lambda_u\, ,
 &\partial_vC&\, = \,\DD_1^G\lambda_v\, ,\\
 \partial_u\lambda_v-\partial_v\lambda_u
 &\, = \, -\DD_2^G\mu_2 \, .
 \end{aligned}
\]
The integration map $I_2$ takes vertex values via evaluation $C_0=C(0,0), C_1=C(1,0), C_2=C(0,1)$ and the parameter values
\[
 \begin{aligned}
 \Lambda_1^{01}&\, := \,\int_0^1\lambda_u(s,0)\,\dd s\, ,\\
 \Lambda_1^{12}&\, := \,\int_0^1(\lambda_v-\lambda_u)(1-s,s)\,\dd s\, ,\\
 \Lambda_1^{02}&\, := \,\int_0^1\lambda_v(0,s)\,\dd s\, ,\\
 \Lambda_2^{012}&\, := \,\int_{\Delta^2}\mu_2\,\dd u\,\dd v \, .
 \end{aligned}
\]
The intuition for the second parameter here is integration of $\dd u \lambda_u + \dd v \lambda_v$ along the $[12]$-edge of the triangle, parametrized by $\gamma_{12}:s\mapsto (1-s,s)$. 

To extract the compatibility relation between these, we apply Stokes’ theorem in the simplex directions, keeping the coefficients as differential forms on \(M\).
Writing the simplex-degree-\(1\) part as
\[
\lambda \, := \, \dd u\,\lambda_u+\dd v\,\lambda_v \, ,
\]the total-form trivialization equation gives
\[
\dd_{\Delta}\lambda
\, = \,
\dd u\wedge\dd v\,
(\partial_u\lambda_v-\partial_v\lambda_u)
\, = \,
-\dd u\wedge\dd v\,\DD_2^G\mu_2 \, .
\] Now given the orientation $\dd u\wedge \dd v$ on the triangle, its oriented boundary is
\[
\partial[012] \, = \, [12]-[02]+[01] \, .
\]Hence Stokes' theorem gives
\[
\begin{aligned}
\Lambda_1^{12}-\Lambda_1^{02}+\Lambda_1^{01}
&\, = \, \int_{\partial\Delta^2}\lambda\\
&\, = \, \int_{\Delta^2}\dd_{\Delta}\lambda\\
&\, = \, -\DD_2^G
\int_{\Delta^2}\mu_2\,\dd u\,\dd v\\
&\, = \, -\DD_2^G\Lambda_2^{012} \, .
\end{aligned}
\]Here we moved \(\DD_2^G\) outside the integral because it acts on the \(M\)-variables, and \(G\) is independent of the simplex coordinates. 
Thus the integrated two-form coefficient is exactly a level-$2$
parameter from the composite $\Lambda_1^{01}+\Lambda_1^{12}$ to
$\Lambda_1^{02}$, a 2-simplex of $\Triv$.

In the other direction, starting with some 2-simplex $(C^a,\Lambda^{ab}_1,\Lambda^{012}_2)_{0\leq a<b \leq2}$ of $T_{\mathrm {gp}}$, let 
\[
 \omega_{ab}\, := \,u_a\dd u_b-u_b\dd u_a\, ,\qquad
 \omega_{012}\, := \,2\dd u\wedge\dd v \, ,
\]
and define the normalized representative
\begin{equation}\label{WGPC:eq:triangle}
 W_2(C,\Lambda)\, := \,\sum_{a=0}^2u_aC_a
 +\omega_{01}\Lambda_1^{01}+\omega_{12}\Lambda_1^{12}
 +\omega_{02}\Lambda_1^{02}+\omega_{012}\Lambda_2^{012} \, .
\end{equation}
To check explicitly this defines a 2-simplex of $\Triv_{\mathrm{total}}$, write $A=\Lambda_1^{01}$,
$B=\Lambda_1^{12}$, $D=\Lambda_1^{02}$, $L=\Lambda_2^{012}$,
and $E=D-A-B=\DD_2^G L$. Then \eqref{WGPC:eq:triangle} expands in the coordinate form basis as
\[
 \begin{aligned}
 W_2(C,\Lambda)\, = \,{}&C_0+u\DD_1^G A+v\DD_1^G D\\
 &+\dd u\,(A+vE)+\dd v\,(D-uE)
                 +2\dd u\wedge\dd v\,L \, .
 \end{aligned}
\]
The one-parameter equations then follow from $\DD_1^G E=0$, and the remaining equation is
\[
 \partial_u(D-uE)-\partial_v(A+vE)+\DD_2^G(2L)
 \, = \, -2E+2\DD_2^G L\, = \,0 \, .
\]
Thus, its edges are the normalized $W_1$ representatives and 
$$
I_2W_2=\mathrm{id} \, .
$$
Note, again, the opposite composite need not recover the original smooth triangle
$$
W_2 I_2 \, \neq \, \mathrm{id}\, ,
$$
Yet $W_2 I_2$ may still be seen to be \textit{homotopic} to the identity, via a homotopy that fixes every triangle in the image of $W_2$.
 
This is made precise and generalized to higher degrees in the Thm. \ref{WGPC:prop:comparison} below, but before that we extend the integration $I$ and the ``Whitney'' W maps to all simplices.
\begin{proposition}[\bf Integration and Whitney reconstruction]\label{IntegrationAndWhitneyRec} There exist natural simplicial maps:
\begin{itemize}
\item Integrating total-forms along simplex faces 
$$
I \, : \, \Triv_{\mathrm{total}}\longrightarrow \Triv \, , 
$$
and constructing the corresponding Whitney form from gauge parameter simplices
$$
W\, : \, \Triv \longrightarrow \Triv_{\mathrm{total}} \, . 
$$
\item In particular 
$$
I\circ W = \mathrm{id}
$$
so that $I$ is level-wise surjective.
\end{itemize}
\end{proposition}

\begin{proof}
\smallskip\emph{The integration map.} We start by defining the integration map on any simplex of $\Triv_{\mathrm{total}}$.
Recall \eqref{TotalFormsEquivalentEquation} that the defining equation from Def. \ref{WGPC:def:forms} is equivalent to
\begin{equation}\label{WGPC:eq:components}
 \DD_0^G\mathcal C_{[0]}\, = \,G\, ,\qquad
 \dd_\Delta\mathcal C_{[r-1]}+(-1)^r\DD_r^G\mathcal C_{[r]}
 \, = \,0\qquad(1\leq r\leq n) \, .
\end{equation}
where $\mathcal C=\sum_{r=0}^n\mathcal C_{[r]}$ is  the decomposition by simplex degree of any such total-form on $M\times \Delta^n$.

Define, coefficient-wise with parameter differentials on the left,
\begin{equation}\label{WGPC:eq:integration}
 (I_n\mathcal C)_r^{a_0\cdots a_r}
 \, := \,\int_{[a_0\cdots a_r]}\mathcal C_{[r]} \, ,
\end{equation}
where for $r=0$ this is simply vertex evaluation. The integral removes only the simplex differentials, leaving a form
in $\mathcal A_r(M)$. Applying Stokes to
\eqref{WGPC:eq:components} gives
\[
 \begin{aligned}
 \sum_{j=0}^r(-1)^j(I_n\mathcal C)_{r-1}^{a_0\cdots\widehat{a_j}\cdots a_r}
  &\, = \,\sum_{j=0}^r (-1)^j \int_{[a_0\cdots \widehat{a_j}\cdots  a_r]}\mathcal C_{[r-1]}
 \\
  &\, = \,\int_{\partial [a_0\cdots a_r]}\mathcal C_{[r-1]}\\
 &\, = \,\int_{[a_0\cdots a_r]}\dd_\Delta\mathcal C_{[r-1]}\\
 &\, = \,(-1)^{r+1}\DD_r^G(I_n\mathcal C)_r^{a_0\cdots a_r} \, .
 \end{aligned}
\]
Thus \eqref{WGPC:eq:integration} satisfies \eqref{WGPC:eq:compatibility}, yielding an $n$-simplex of $\Triv$.

\smallskip\emph{The Whitney reconstruction map.} For the converse direction, recall the ``Whitney form'' of an ordered face is
\[
 \omega_{a_0\cdots a_r}\, := \,r!\sum_{j=0}^r(-1)^j u_{a_j}
 \dd u_{a_0}\wedge\cdots\wedge\widehat{\dd u_{a_j}}
                  \wedge\cdots\wedge\dd u_{a_r} \, ,
\]
where in degree zero this means simply $\omega_a=u_a$. For any $n$-simplex of gauge parameters
\[
 (C,\Lambda)
 \, = \,
 \bigl(\Lambda_r^{a_0\cdots a_r}\bigr)_
 {\substack{0\leq r\leq n\\0\leq a_0<\cdots<a_r\leq n}}
 \, ,\qquad
 \Lambda_0^a\, := \,C_a \, ,
\]
define
\begin{equation}\label{WGPC:eq:Whitney}
 W_n(C,\Lambda)\, := \,\sum_{r=0}^n
 \sum_{a_0<\cdots<a_r}\omega_{a_0\cdots a_r}
                       \Lambda_r^{a_0\cdots a_r}\, ,
 \qquad \Lambda_0^a\, := \,C_a \, .
\end{equation}
A small combinatorial calculation using the barycentric coordinate identity shows that Whitney reconstruction intertwines simplicial coboundary and simplex
exterior derivative \cite[Prop.~3.4(3)]{Lunardon}:
\[
 \dd_\Delta W(\Lambda_{r-1})\, = \,W(\delta\Lambda_{r-1}) \, ,
\]
where recall
\[
(\delta\Lambda_{r-1})^{a_0\cdots a_r}
\, := \,
\sum_{j=0}^r(-1)^j
\Lambda_{r-1}^{a_0\cdots\widehat{a_j}\cdots a_r} \, .
\]
It also commutes with the coefficient operators $\DD_r^G$, since these only act on the  $M$-form coefficients.
Consequently \eqref{WGPC:eq:compatibility} implies every positive-degree
equation in \eqref{WGPC:eq:components}
\[
 \begin{aligned}
 \dd_\Delta W_n(\Lambda_{r-1})
       +(-1)^r\DD_r^G W_n(\Lambda_r)
 &\, = \,W_n\bigl(\delta\Lambda_{r-1}
                       +(-1)^r\DD_r^G\Lambda_r\bigr)\\
 &\, = \,0 \, .
 \end{aligned}
\]
The degree-zero equation follows directly from
\[
 \DD_0^G\left(\sum_a u_aC_a\right)
 \, = \,\sum_a u_aG\, = \,G \, ,
\]
and hence \eqref{WGPC:eq:Whitney} is indeed a total-form n-simplex of $T_\mathrm{forms}$.

\smallskip\emph{Normalization.}
For faces of the same dimension, the normalization of Whitney forms is exactly such that
\[
 \int_{[b_0\cdots b_r]}\omega_{a_0\cdots a_r}
 \, = \,\begin{cases}1,&(a_0,\ldots,a_r)=(b_0,\ldots,b_r),\\
 0,&\text{otherwise}\, ,\end{cases}
\]
which proves \cite[proof of Thm.~4.3(1)]{Lunardon}
$$
IW=\mathrm{id}\, .
$$
Indeed, starting with a gauge-parameter simplex \(\Lambda\), its reconstructed simplex-degree-\(r\) component is
\[
\bigl(W_n\Lambda\bigr)_{[r]}
\, = \,
\sum_{a_0<\cdots<a_r}
\omega_{a_0\cdots a_r}\Lambda_r^{a_0\cdots a_r} \, .
\]Integrating this over a particular face gives
\[
\begin{aligned}
\bigl(I_nW_n\Lambda\bigr)_r^{b_0\cdots b_r}
&\, = \,
\sum_{a_0<\cdots<a_r}
\left(
\int_{[b_0\cdots b_r]}\omega_{a_0\cdots a_r}
\right)\Lambda_r^{a_0\cdots a_r}\\
&\, = \,
\Lambda_r^{b_0\cdots b_r} \, .
\end{aligned}
\]

\smallskip\emph{Simplicial naturality.}
So far, we have shown \(I_n\) and \(W_n\) yield a collection of maps in each dimension. To assemble them into maps of simplicial sets, they must satisfy:
For every monotone map \(\theta:[m]\to[n]\)
\[
I_m(\theta^*\mathcal C)
\, = \,
\theta^*(I_n\mathcal C),
\qquad
W_m(\theta^*\Lambda)
\, = \,
\theta^*(W_n\Lambda) \, .
\]
Consider an \(r\)-face \([a_0\cdots a_r]\) of the source simplex.
If its vertices have distinct images under \(\theta\), the associated affine map sends this face bijectively, preserving orientation, onto
\([\theta(a_0)\cdots\theta(a_r)]\). Therefore change of variables gives
\[
\int_{[a_0\cdots a_r]}\theta^*\mathcal C_{[r]}
\, = \,
\int_{[\theta(a_0)\cdots\theta(a_r)]}\mathcal C_{[r]} \, ,
\]which is exactly the parameter assigned by \(\theta^*I_n\mathcal C\).
If some vertices have the same image, the face collapses to dimension less than \(r\), and its pullback of an \(r\)-form vanishes, and hence both sides are zero.

Naturality of $W$ follows by the relevant Whitney pullback formula
\[
(\theta_*)^*\omega_{b_0\cdots b_r}
\, = \,
\sum_{\substack{a_0<\cdots<a_r\\
\theta(a_j)=b_j}}
\omega_{a_0\cdots a_r} \, , 
\]
which is exactly \cite[Prop 3.4(2)]{Lunardon}, translated into our indices. Indeed, 
\[
\begin{aligned}
\theta^*W_n(\Lambda_r)
&\, = \,
\sum_{b_0<\cdots<b_r}
(\theta_*)^*\omega_{b_0\cdots b_r}\,
\Lambda_r^{b_0\cdots b_r}\\
&\, = \,
\sum_{b_0<\cdots<b_r}
\sum_{\substack{a_0<\cdots<a_r\\\theta(a_j)=b_j}}
\omega_{a_0\cdots a_r}\,
\Lambda_r^{b_0\cdots b_r}\\
&\, = \,
\sum_{\substack{a_0<\cdots<a_r\\
\theta(a_0)<\cdots<\theta(a_r)}}
\omega_{a_0\cdots a_r}\,
\Lambda_r^{\theta(a_0)\cdots\theta(a_r)} \, .
\end{aligned}
\]
But our definition is precisely
\[
(\theta^*\Lambda)_r^{a_0\cdots a_r}
\, = \,
\begin{cases}
\Lambda_r^{\theta(a_0)\cdots\theta(a_r)},
&\text{if the images are distinct},\\
0,&\text{otherwise}.
\end{cases}
\]
We can therefore include the collapsed faces as zero summands. Thus
\[
\theta^*W_n(\Lambda_r)
\, = \,
\sum_{a_0<\cdots<a_r}
\omega_{a_0\cdots a_r}
(\theta^*\Lambda)_r^{a_0\cdots a_r}
\, = \,
W_m(\theta^*\Lambda_r) \, ,
\] i.e., 
\[
\theta^*W_n(\Lambda)
\, = \,
W_m(\theta^*\Lambda) \, .
\]
\end{proof}

\begin{theorem}[\bf Comparison of weighted trivialization models]
\label{WGPC:prop:comparison}
Under Assumption~\ref{ass:main}, let $G$ be a curvature on a smooth manifold $M$.
\begin{enumerate}
\item The simplicial maps of Proposition~\ref{IntegrationAndWhitneyRec}
\[
 \Triv_{\mathrm{total}}
 \ \underset{W}{\overset{I}{\rightleftarrows}}\
 \Triv
\]
satisfy $IW=\mathrm{id}$ and admit a simplicial homotopy
\[
 H:\Triv_{\mathrm{total}}\times\Delta[1]\longrightarrow\Triv_{\mathrm{total}}\,,
 \qquad H|_0\,=\,\mathrm{id}\,,
 \qquad H|_1\,=\,WI\,,
\]
which fixes $\operatorname{im}(W)$.

\item Both simplicial sets are Kan, and hence these maps exhibit an equivalence
of infinity groupoids
\[
 \Triv_{\mathrm{total}}\,\simeq\,\Triv\,.
\]
In particular, $W$ identifies $\Triv$ with the Whitney submodel
$T_{\mathrm{Whitney}}:=\operatorname{im}(W)$, which is a strong simplicial
deformation retract of $\Triv_{\mathrm{total}}$.
\end{enumerate}
\end{theorem}
\begin{proof}
Proposition~\ref{IntegrationAndWhitneyRec} already gives simplicial
maps $I,W$ with $IW=\mathrm{id}$. It remains to construct a simplicial
homotopy from $\mathrm{id}$ to $WI$ and check that it fixes Whitney
representatives. If there is no vertex, both simplicial sets are empty.
Otherwise choose a trivialization $C_*$ and suppress its constant
pullback to all parameter spaces below. Put 
$$
a=\mathcal C-C_*
$$ with
$P_n=W_nI_n$ acting on these differences.

\smallskip\noindent\emph{Strategy.}
We seek a form $\mathcal H_n(\mathcal C)$ on
$M\times\Delta^n\times I_s$ satisfying
\[
 \dd_{M\times\Delta^n\times I_s}\mathcal H_n(\mathcal C)
 +R(\mathcal H_n(\mathcal C),G)
 \, = \, G \, ,
\]
with $I_s$-endpoint restrictions $\mathcal C$ and $W_nI_n\mathcal C$,
naturally in the simplex, so that pullback along the barycentric
graphs will define a \textit{simplicial} homotopy
$H:\Triv_{\mathrm{total}}\times\Delta[1]\to \Triv_{\mathrm{total}}$ (see e.g. \cite[Definition~14.26.1]{Stacks}). The problem with the naive scalar interpolation
$$
C_*+(1-s)a+sP_na
$$
is that, although it
has
the required endpoints and solves the equation for each fixed $s$, its exterior differential on the product has an additional $\dd s$
term. Lemma~\ref{WGPC:lem:naturalprimitive} below supplies a primitive
that cancels this defect to create $\mathcal{H}_n$. We carry out this construction in three steps.

\smallskip\noindent\emph{Step 1: construct a total-form family on an interval.}
Write $B_n=M\times\Delta^n$ and introduce the interval variable $s\in I_s$. Set
\[
 A_s\, := \,(1-s)a+sP_na \, ,\qquad
 \mathcal C_s\, := \,C_*+A_s \, ,
\]
where both $a$ and $P_n a$ satisfy the homogeneous trivialization equation,
so for each fixed $s$,
\[
 \dd_{B_n}A_s+R(A_s,G)\, = \,0 \, ,\qquad
 \dd_{B_n}\mathcal C_s+R(\mathcal C_s,G)\, = \,G \, .
\]
Notice here
$
P_nC_*=W_nI_n(C_*)
=
C_*$, since 
\[
 (I_nC_*)_0^a\, = \,C_* \, ,\qquad
 (I_nC_*)_r^{a_0\cdots a_r}\, = \,0\quad(r\geq1) \, ,\qquad
 W_nI_nC_*\, = \,\sum_a u_aC_*\, = \,C_* \, ,
\]
as  $C_*$ has no differential legs along $\Delta^n$, and by the barycentric coordinate identity $\sum_a u_a=1$.
When $\mathcal C_s$ is viewed as a total form on the product with the interval $I_s$, however,
\[
 \dd_{B_n\times I_s}\mathcal C_s+R(\mathcal C_s,G)-G
 \, = \,\dd s\wedge(P_na-a) \, ,
\]
since the component differential $\dd s \wedge \partial_s$ acts on $A_s$ non-trivially (cf. \eqref{1SimplexTotalFailure} for the 1-simplex case.)

Nevertheless, by Lemma~\ref{WGPC:lem:naturalprimitive} below, there is a simplicially
natural collection $b_n(a)$, whose $i$-th component has degree $g^i-2$,
such that
\begin{equation}\label{WGPC:eq:primitiveforms}
 \dd_{B_n}b_n(a)-R(b_n(a),G)
 \, = \,a-P_na \, .
\end{equation}
Thus, define
\begin{align*}
 \mathcal H_n(\mathcal C)
 \, :&= \,\mathcal C_s-\dd s\wedge b_n(a)
\\
&= \,C_*+(1-s)a+sP_na-\dd s\wedge b_n(a) \, ,
\end{align*}
and note that since $b_n(a)$ is independent of $s$,
\[
 \begin{aligned}
 \dd_{B_n\times I_s}(-\dd s\wedge b_n(a))
 &\, = \,\dd s\wedge\dd_{B_n}b_n(a) \, ,\\
 R(-\dd s\wedge b_n(a),G)
 &\, = \,-\dd s\wedge R(b_n(a),G) \, .
 \end{aligned}
\]
Consequently, by \eqref{WGPC:eq:primitiveforms},
\[
 \begin{aligned}
 &\dd_{B_n\times I_s}\mathcal H_n(\mathcal C)
       +R(\mathcal H_n(\mathcal C),G)-G\\
 &\qquad\, = \,\dd s\wedge
 \bigl(P_na-a+\dd_{B_n}b_n(a)-R(b_n(a),G)\bigr)
 \, = \,0 \, .
 \end{aligned}
\]
This verifies the full trivialization equation for $\mathcal H_n$ on the product, including the
$\dd s$ component. Since restricting to an endpoint kills $\dd s$, we also have
\[
 \mathcal H_n(\mathcal C)|_{s=0}\, = \,\mathcal C \, ,\qquad
 \mathcal H_n(\mathcal C)|_{s=1}\, = \,W_nI_n\mathcal C \, .
\]

\smallskip\noindent\emph{Step 2: obtain a simplicial homotopy.}
The $I_s$-interval family must now define a map
$H:\Triv_{\mathrm{total}}\times\Delta[1]\to \Triv_{\mathrm{total}}$.
For an $n$-simplex of $\Delta[1]$, i.e., a monotone map $\alpha:[n]\to[1]$,
associate to it the affine function $s_\alpha : \Delta^n \rightarrow \Delta^1\cong I_s$ and its graph $\Gamma_\alpha : M\times \Delta^n \rightarrow M\times \Delta^n \times I_s$
\[
 s_\alpha(u)\, := \,\sum_{\alpha(j)=1}u_j \, ,\qquad
 \Gamma_\alpha(m,u)\, := \,\bigl(m,u,s_\alpha(u)\bigr) \, ,
\]
and use it to define 
\[
 \begin{aligned}
 H_n(\mathcal C,\alpha)
 &\, := \,\Gamma_\alpha^*\mathcal H_n(\mathcal C)\\
 &\, = \,C_*+(1-s_\alpha)a+s_\alpha P_na
                       -\dd s_\alpha\wedge b_n(a) \, .
 \end{aligned}
\]
Pullback of the total family preserves the trivialization equation,
so this is indeed an $n$-simplex of $\Triv_{\mathrm{total}}$.
For a monotone map $\theta:[m]\to[n]$, writing $v_j$ for the source barycentric
coordinates, we have
\[
 \theta^* s_\alpha \,=\,  s_\alpha\circ\theta_*
 \, = \,\sum_{\alpha(i)=1}\sum_{\theta(j)=i}v_j
 \, = \,\sum_{\alpha(\theta(j))=1}v_j
 \, = \,s_{\alpha\circ\theta} \, .
\]
Together with $\theta^*P_n=P_m\theta^*$, $\theta^* a = \theta^* \mathcal{C} - C_* $ and the naturality
$\theta^*b_n(a)=b_m(\theta^*a)$ from
Lemma~\ref{WGPC:lem:naturalprimitive}, substitution gives
\[
\begin{aligned}
\theta^*H_n(\mathcal C,\alpha)
&\,=\,
\theta^*\!\left(
C_*+(1-s_\alpha)a+s_\alpha P_na
-\dd s_\alpha\wedge b_n(a)
\right)\\
&\,=\,
C_*+(1-\theta^*s_\alpha)\theta^*a
+(\theta^*s_\alpha)\theta^*(P_na)
-\theta^*(\dd s_\alpha)\wedge\theta^*b_n(a)\\
&\,=\,
H_m(\theta^*\mathcal C,\alpha\circ\theta) \, .
\end{aligned}
\]
This verifies all simplicial operators, including degeneracies.
If $\alpha$ is constantly zero or one, then $s_\alpha$ is respectively
zero or one and $\dd s_\alpha=0$, so that the two restrictions of $H$ are
therefore $\mathrm{id}$ and $WI$.

\smallskip\noindent\emph{Step 3: check the fixed submodel and conclude.}
If $\mathcal C$ is a Whitney representative, $\mathcal{C}=P_n \mathcal{C}$, then
$a=\mathcal C-C_*$ is also Whitney. Thus $P_na=a$ and
$b_n(a)=0$ by Lemma~\ref{WGPC:lem:naturalprimitive}, and hence
\[
 H_n(\mathcal C,\alpha)\, = \,\mathcal C
 \qquad\text{for every }\alpha:[n]\to[1] \, .
\]
This proves the first assertion.

For the second, combine this homotopy
with $IW=\mathrm{id}$ from Proposition~\ref{IntegrationAndWhitneyRec}.
Both simplicial sets are Kan by affine linearity after translation
by $C_*$, so the resulting simplicial homotopy equivalence is an
equivalence of infinity groupoids. Finally, $IW=\mathrm{id}$ makes
$W$ injective in every simplicial degree, with inverse $I$ on its image, giving
the stated isomorphism $\Triv\cong T_{\mathrm{Whitney}}$.
\end{proof}

\begin{lemma}[\bf Natural primitives for Whitney reconstruction]
\label{WGPC:lem:naturalprimitive}
Under the preceding hypotheses, let $a$ be a collection of forms on
$M\times\Delta^n$, with $a^{(i)}$ of degree $g^i-1$, satisfying
\[
 \dd_{M\times\Delta^n}a+R(a,G)\, = \,0 \, .
\]
There is a linear, simplicially natural choice of forms $b_n(a)$,
with $b_n(a)^{(i)}$ of degree $g^i-2$, satisfying
\eqref{WGPC:eq:primitiveforms}, where $P_n=W_nI_n$ acts on the
homogeneous form collections. Moreover, $b_n(a)=0$ whenever $P_na=a$.
\end{lemma}
\begin{proof}
\smallskip\noindent\emph{Step 1: translate homogeneous solutions into cocycles.}
For $(L^\bullet,\partial_L)$ the gauge complex of
Def.~\ref{WGPC:def:gaugecomplex}, denote by
$\Omega^p(\Delta^n;L^q)$ the simplex $p$-forms whose coefficients
are smooth families of the $M$-forms in $L^q$. In particular, for
$q=0$ their coefficients lie in $\ker\DD_0^G$ at every parameter
value. Define the total complex by
\begin{align}\label{TotalGaugeComplex}
 \mathcal E_n^k\, := \,
 \bigoplus_{p=0}^n\Omega^p(\Delta^n;L^{k-p}) \, ,\qquad
 \mathcal D(\alpha\otimes z)
 \, := \,\dd_\Delta\alpha\otimes z
       +(-1)^p\alpha\otimes\partial_Lz
 \qquad(\deg\alpha=p) \, ,
\end{align}
where the displayed differential specifies the convention on separated terms,
while for general smooth families $\dd_\Delta$ also differentiates the
parameter dependence of the coefficients. In particular, no finite-sum tensor-product
restriction is imposed. The two component differentials square to zero
and anticommute after the displayed sign, so $\mathcal D^2=0$.

The component $a_{[r]}$ of $a \in\prod_i\Omega^{g^i-1}(M\times\Delta^n)$ has simplex differential degree $r$ and
gauge coefficient degree $-r$ in $L^{-r} =\prod_i\Omega^{g^i-1-r}(M)$ for $r\geq 1$. In other words, $a\in\mathcal E_n^0$  where
\[
 \mathcal E_n^0
 \, = \,\Omega^0(\Delta^n;L^0)
 \oplus\Omega^1(\Delta^n;L^{-1})
 \oplus\Omega^2(\Delta^n;L^{-2})\oplus\cdots \, .
\]
The degree-zero coefficient lies in $L^0$ because
$\DD_0^G a_{[0]}=0$, while the remaining homogeneous versions of
\eqref{WGPC:eq:components} are
\[
 \dd_\Delta a_{[r-1]}+(-1)^r\DD_r^Ga_{[r]}\, = \,0
 \qquad(1\leq r\leq n) \, .
\]
But these are exactly the components of $\mathcal D a=0$ from \eqref{TotalGaugeComplex}, so that,
when a vertex $C_*$ exists, translation identifies
\begin{align}\label{nSimplexAsCocycle}
 (\Triv_{\mathrm{total}})_n-C_*
 \, = \,Z^0(\mathcal E_n^\bullet) \, .
\end{align}
The identification \eqref{nSimplexAsCocycle} explains why a contraction
of the cochain complex \eqref{TotalGaugeComplex} supplies the required primitives.

\smallskip\noindent\emph{Step 2: identify the Dupont contraction being used.}
On scalar simplex forms $\Omega^\bullet(\Delta^n)$, write $p_n$ for
integration over faces followed by Whitney reconstruction:
\[
 p_n\eta\, := \,\sum_{r=0}^n\sum_{a_0<\cdots<a_r}
 \omega_{a_0\cdots a_r}
 \int_{[a_0\cdots a_r]}\eta_{[r]} \, .
\]
In Lunardon's \cite{Lunardon} notation, $C_n$ is the space of Whitney
forms, $\pi_n:\Omega^\bullet(\Delta^n)\to C_n$ is this reconstruction
with codomain restricted to that space, and $i_n$ is its inclusion,
thus $p_n=i_n\pi_n$. Theorem~3.8 of \cite{Lunardon} supplies the
simplicial contraction. With $h_n$ the negative of the Dupont homotopy operator in that
source (originally due to \cite[Thm. 2.16]{Dupont78}), its identities are
\begin{align}\label{ScalarDupontIdentity}
 \dd_\Delta h_n+h_n\dd_\Delta\, = \,\mathrm{id}-p_n \, ,\qquad
 (\theta_*)^*h_n\, = \,h_m(\theta_*)^*
 \quad(\theta:[m]\to[n]) \, .
\end{align}
We also use $h_np_n=0$, equivalently that $h_n$ annihilates Whitney
forms (see \cite[Lem.~3.4 and Thm.~3.11]{Getzler}).
These identities concern the simplex exterior differential
$\dd_\Delta$, before introducing our coefficient differential.

The cited constructions in \cite{Lunardon,Getzler} use polynomial
forms. For the smooth-form contraction and its side conditions, see
\cite[Thms.~2.3.1--2.3.2]{Albert}, whose operators $R,s$ correspond
to our scalar integration and $h_n$. The formulas are finite sums of
multiplication by Whitney forms, pullbacks along straight-line
contractions, and integration over compact parameter intervals. They
therefore act on forms smooth up to the boundary and on smooth
$M$-dependent families. The same pullback and integration identities
give simplicial naturality, including degeneracies, in this setting. Notably, this does not require a finite algebraic tensor-product decomposition.

\smallskip\noindent\emph{Step 3: extend the Dupont contraction to the gauge complex.}
Extend $p_n,h_n$ to the total gauge complex \eqref{TotalGaugeComplex} by acting only in the
simplex directions, leaving the $M$-form coefficients untouched. For the analogous comparison of smooth simplicial forms with manifold-form coefficients, see \cite[
Thm. 6.4]{Dupont78}. Here we verify compatibility with the gauge differential coefficientwise.
Denote the resulting projection by $P_n$, so that on translated
trivializations it is $W_nI_n$. The operator $h_n$ lowers simplex form
degree by one and preserves coefficient degree, so
\[
 h_n:\mathcal E_n^k\longrightarrow\mathcal E_n^{k-1} \, .
\]
For $\deg\alpha=p$, the two contributions involving $\partial_L$
in the total contraction calculation cancel:
\[
 \begin{aligned}
 \mathcal D h_n(\alpha\otimes z)
 &\, = \,\dd_\Delta h_n\alpha\otimes z
       +(-1)^{p-1}h_n\alpha\otimes\partial_Lz \, ,\\
 h_n\mathcal D(\alpha\otimes z)
 &\, = \,h_n\dd_\Delta\alpha\otimes z
       +(-1)^p h_n\alpha\otimes\partial_Lz \, ,
 \end{aligned}
\]
while for $p=0$ the terms containing $h_n\alpha$ vanish. The same cancellation
holds for general smooth families, since the parameter operators
commute with $\partial_L$. Hence \eqref{ScalarDupontIdentity} implies the analogous contraction identities on the total complex
\[
 \mathcal D h_n+h_n\mathcal D\, = \,\mathrm{id}-P_n \, ,\qquad
 h_nP_n\, = \,0 \, ,
\]
with $P_n$ and $h_n$ retaining their simplicial naturality.

Now since a homogeneous trivialization $a$ satisfies $\mathcal D a=0$, we may define 
$$
b_n(a)=h_na\in\mathcal E_n^{-1}\, ,
$$
so that, in particular,
\[
 \mathcal D b_n(a)\, = \,a-P_na \, .
\]
In other words, $h_n a$ is a primitive of $a-P_n a$.

To express this identity in terms of forms on $M\times\Delta^n$, a simplex-degree-$p$
component of $b_n(a)$ has gauge coefficient degree $-p-1$, valued in $L^{-1-p}$, and hence
$M$-form degree $g^i-2-p$ in component $i$, since $L^{-1-p} =\prod_i\Omega^{g^i-2-p}(M)$. Its full form degree on $M\times \Delta^n$ is
therefore $g^i-2$, as expected. Since
$\DD_{p+1}^G=\dd_M+(-1)^{p+1}R(-,G)$, the coefficient part of
$\mathcal D$ on this component is
\[
(-1)^p\partial_L \vert_{L^{-1-p}}\,  = \,  (-1)^p\DD_{p+1}^G
 \, = \,(-1)^p\dd_M-R(-,G) \, .
\]
Thus, with simplex differentials on the left this says
\[
 \mathcal D b_n(a)
 \, = \,\dd_{M\times \Delta^n}b_n(a)-R(b_n(a),G) \, ,
\]
which proves \eqref{WGPC:eq:primitiveforms}, since on separated terms $\alpha\wedge z$ of this type
\[
 \begin{aligned}
 \mathcal D(\alpha\otimes z)
 &\, = \,\dd_\Delta\alpha\otimes z
       +(-1)^p\alpha\otimes\DD_{p+1}^Gz\\
 &\, = \,\dd_\Delta\alpha\otimes z
       +(-1)^p\alpha\otimes\dd_Mz
       -\alpha\otimes R(z,G) \, .
 \end{aligned}
\]

Linearity and simplicial
naturality follow from those of $h_n$. Finally, if $P_na=a$, then
$b_n(a)=h_nP_na=0$. Note the same contraction identity underlies the identification of the abelian Dupont-gauge model with the
Dold–Kan model in
\cite[Prop.~5.1]{Getzler}.
\end{proof}

\section{Equivalence to null-concordances}
\label{NullToTrivSubsection}

Let $M$ be a smooth manifold, retain the standing weighted Bianchi algebra
assumption, and fix a curvature $G$ on $M$.
A \textit{null-concordance} of $G$ is a DGCA map from $A= \mathrm{CE}(\frg)$ to differential forms on $M\times I_t$ 
$$
\widehat{G} \, : \, \mathrm{CE}(\frg) \longrightarrow \Omega^\bullet(M\times I_t)
$$
that is equal to $G$ at $t=1$ and $0$ at $t=0$, hence equivalent to a tuple of differential forms of the form
\begin{equation}\label{WNC:eq:hatG}
 \widehat G^{(i)} \, = \, G^{(i)}(t)+\dd t\,\overline C^{(i)}(t) \, ,
 \quad G^{(i)}(0) \, = \, 0 \, ,\quad G^{(i)}(1) \, = \, G^{(i)} \, ,
\end{equation}
which satisfy $\dd_{M\times I_t} \widehat{G}^{(i)} = P^{(i)}(\widehat{G})$. We denote the set of null-concordances of the curvature $G$ on $M$ by
$$
\mathrm{Null}(G)(M)\, . 
$$

The
Bianchi identity with respect to the product differential splits into 
$$
\dd_MG^{(i)}(t)=P^{(i)}(G(t))
$$ 
and the flow equation
\begin{equation}\label{WNC:eq:flow}
 \partial_tG^{(i)} \, = \, \dd_M\overline C^{(i)}
                  +\sum_j\overline C^{(j)}M^{(i)}{}_{(j)}(G(t)) \, .
\end{equation}
Here $\overline C(t)$ is part of the prescribed null-concordance,
whereas $C(t)$ below is the family of potentials to be constructed.
Integrating the flow equation gives only an inductive integral formula for the solution in terms of $G$ in terms of $\bar{C}$ 
\[
 G^{(i)}(t) \, = \, \dd_M\int_0^t\overline C^{(i)}(u)\,\dd u
 +\int_0^t\sum_j\overline C^{(j)}(u)M^{(i)}{}_{(j)}(G(u))\,\dd u \, ,
\]
whose evaluation at $t=1$ is generally \textit{not} of the weighted trivialization form from Eq. \eqref{TrivializingGi}, and in particular \textit{not linear} in the would-be gauge parameter $\int_0^t\overline C^{(i)}(u)\,\dd u$. Nevertheless, the following correction shows that this expression may be massaged into the prescribed
weighted trivialization form.

\begin{proposition}[\bf Weighted projection]\label{WNC:prop:projection}
For any null-concordance \eqref{WNC:eq:hatG}, define the one-parameter family of gauge parameters $C^{(i)}(t)$ in
increasing weight by $C^{(i)}(0)=0$ and the ODE
\begin{align}
 \partial_tC^{(i)}(t)& \, = \, \overline C^{(i)}(t)+X^{(i)}(t) \, ,
 \label{WNC:eq:recursiveC}
 \end{align}
where the correction $X^{(i)}(t)$ is defined recursively by lower weight variables
 \begin{align}
 X^{(i)}(t)
 & \, := \, \frac1{w_i}\sum_{j,k}w_j(-1)^{g^j}
 C^{(j)}(t)\,\overline C^{(k)}(t)\,
 \partial_kM^{(i)}{}_{(j)}(G(t)) \, . \label{WNC:eq:X}
\end{align}
Then:
\begin{itemize}
\item The $1$-parameter family of curvatures $G^{(i)}(t)$ satisfies the weighted trivialization formula w.r.t $C^{(i)}(t)$
\begin{equation}\label{WNC:eq:trivatT}
 G^{(i)}(t) \, = \, \dd_MC^{(i)}(t)+R^{(i)}(C(t),G(t)) \, .
\end{equation}
\item In particular, evaluation at $t=1$ yields a well-defined map from null concordances to weighted trivializations
\begin{align*}
 P_W\,:\, \Null(G)(M)&\xrightarrow{\quad \quad }\Triv_W(G)(M) \equiv (\Triv)_0 \\
 G^{(i)}(t) +\dd t \overline{C}^{(i)}(t) \, &\quad \longmapsto \quad\, C^{(i)}(1) \, ,
\end{align*}
natural under pullback in $M$.
\end{itemize}
\end{proposition}
\begin{proof}
We prove inductively the stronger identity on the whole product,
\[
 (\dd_M+\dd_t)C^{(j)}(t)
 \, = \, \widehat G^{(j)}-R^{(j)}(C(t),\widehat G) \, .
\]
At the lowest weights $P^{(j)}=0$, and this follows by integrating
the flow equation with zero initial value. Suppose this identity
holds at all weights smaller than $w_i$ and define
\[
 Y^{(i)} \, := \, \widehat G^{(i)}-
\frac1{w_i}\sum_jw_jC^{(j)}(t)
M^{(i)}{}_{(j)}(\widehat G) \, .
\]
Here $C^{(j)}(t)$ is viewed as a form on $M\times I$ with no
$\dd t$-component. To compute its total differential, we abbreviate
$\dd_{M\times I}=\dd_M+\dd_t$ and suppress the argument
$\widehat G$ in $P$ and $M^{(i)}{}_{(j)}$ in the following calculation.
Using the induction hypothesis and the graded Leibniz rule gives
\begin{align*}
 \dd_{M\times I}Y^{(i)}
 &\, = \,P^{(i)}
 -\frac1{w_i}\sum_jw_j
 \left(\dd_{M\times I}C^{(j)}\,M^{(i)}{}_{(j)}
 +(-1)^{g^j-1}C^{(j)}\dd_{M\times I}M^{(i)}{}_{(j)}\right)\\
 &\, = \,P^{(i)}
 -\frac1{w_i}\sum_jw_j\widehat G^{(j)}M^{(i)}{}_{(j)}
 +\frac1{w_i}\sum_{j,k}w_k C^{(k)}
                   M^{(j)}{}_{(k)}M^{(i)}{}_{(j)}\\
 &\hspace{1.2em}
 -\frac1{w_i}\sum_k w_k(-1)^{g^k-1}C^{(k)}
                         \dd_{M\times I}M^{(i)}{}_{(k)}\\
 &\, = \,\frac1{w_i}\sum_k w_k C^{(k)}
 \left(\sum_jM^{(j)}{}_{(k)}M^{(i)}{}_{(j)}
       +(-1)^{g^k}\dd_{M\times I}M^{(i)}{}_{(k)}\right)
 \, = \,0 \, ,
\end{align*}
where in the second equality we substituted
$\dd_{M\times I}C^{(j)}=\widehat G^{(j)}-R^{(j)}(C,\widehat G)$
and expanded $R^{(j)}$. The weighted Euler identity cancels the
terms $P^{(i)}$ and $-w_i^{-1}\sum_jw_j\widehat G^{(j)}M^{(i)}{}_{(j)}$, and the
Bianchi derivative identity of Lem \ref{lem:matrix} makes each parenthesis
in the last line vanish.

Next, expanding $Y^{(i)}=Y^{(i)}_0+\dd t\,Y^{(i)}_1$, we have
\begin{align*}
Y^{(i)}_0+\dd t\,Y^{(i)}_1 \, & = \, \widehat G^{(i)}- \frac1{w_i}\sum_jw_jC^{(j)}(t)   M^{(i)}{}_{(j)}(\widehat G) \, , 
\end{align*}
where by plugging in $\widehat G^{(i)} \, = \, G^{(i)}+\dd t\,\overline C^{(i)}$ this yields
\begin{align*}
Y^{(i)}_0+\dd t\,Y^{(i)}_1 \, & = \, G^{(i)}+\dd t\,\overline C^{(i)} - \frac1{w_i}\sum_jw_jC^{(j)}   M^{(i)}{}_{(j)}(G) - \frac1{w_i}\sum_{j,k}w_jC^{(j)} \dd t\,  \overline C^{(k)}\partial_kM^{(i)}{}_{(j)}(G) \, .
\end{align*}
Moving $\dd t$ past $C^{(j)}$ contributes
$(-1)^{g^j-1}$,  therefore
\[
 Y^{(i)}_0 \, = \, G^{(i)}-R^{(i)}(C,G) \qquad  ,\qquad
 Y^{(i)}_1 \, = \, \overline C^{(i)}+X^{(i)} \, .
\]
and the proven closure of $Y^{(i)}$ implies that $\partial_tY^{(i)}_0=\dd_MY^{(i)}_1$, while also $Y^{(i)}_0(0)=0$. Hence, setting
$C^{(i)}(t)=\int_0^tY^{(i)}_1(u)\,\dd u$ gives
$$
\dd_MC^{(i)}(t)\, =\, Y^{(i)}_0(t)\, = \, G^{(i)}-R^{(i)}(C,G)
$$ and further
$(\dd_M+\dd_t)C^{(i)}(t)=Y^{(i)}$, which proves both the induction and \eqref{WNC:eq:trivatT}.
All operations commute with pullback in $M$.
\end{proof}

\begin{example}[\bf $\mathfrak{l}S^4$-weighted projection]
Consider $\dd G^{(1)}=0$ and $\dd G^{(2)}=\frac12(G^{(1)})^2$ with $g^1=4$ and $g^2=7$,
$w_2=2w_1$, and let
$B^{(i)}(t)=\int_0^t\overline C^{(i)}(u)\,\dd u$. Then
\[
 C^{(1)}(t) \, = \, B^{(1)}(t) \, ,\qquad
 X^{(2)}(t) \, = \, \frac12B^{(1)}(t)\overline C^{(1)}(t)
 \, = \, -\frac12\overline C^{(1)}(t)B^{(1)}(t) \, .
\]
Thus
\[
 C^{(2)}(t) \, = \, B^{(2)}(t)-\frac12\int_0^t
                 \partial_uB^{(1)}(u)B^{(1)}(u)\,\dd u \, ,
\]
and the resulting 1-parameter trivialization is, as expected, 
\[
 G^{(2)}(t) \, = \, \dd_M C^{(2)}(t)
       +\frac12 C^{(1)}(t)G^{(1)}(t) \, .
\]
\end{example}

Accordingly, from any weighted trivialization one may reconstruct a  null-concordance by assigning the appropriate $t$-interval dependence in a weighted manner.

\begin{proposition}[\bf Weighted section]
\label{WNC:prop:section}
For any weighted trivialization $C$ of $G$, put
\begin{equation*}
 \sigma_W(C)^{(i)} \, := \, t^{w_i}G^{(i)}
               +w_it^{w_i-1}\dd t\,C^{(i)} \, .
\end{equation*}
\begin{itemize}
\item
This is a smooth null-concordance for $G$ and hence defines a map
$$
\sigma_W\,:\, (\Triv)_0 \equiv \Triv_W(G)(M) \longrightarrow \Null(G)(M) \, .
$$
\item
Moreover,
$$
P_W\sigma_W=\mathrm{id} \, : \, \Triv_W(G)(M) \longrightarrow \Triv_W(G)(M) \, ,
$$ and so $P_W$ is surjective.
\end{itemize}
\end{proposition}
\begin{proof}
First we verify that the displayed form is a null-concordance.
Write its two components as
\[
 G_t^{(i)} \, := \, t^{w_i}G^{(i)},
 \qquad \overline C_t^{(i)} \, := \, w_it^{w_i-1}C^{(i)} \, ,
\]
and note that since the weights are positive integers, both expressions are smooth
at $t=0$. By weight homogeneity of the Bianchi polynomials, we have
\[
 P^{(i)}(G_t) \, = \, t^{w_i}P^{(i)}(G),
 \qquad
 M^{(i)}{}_{(j)}(G_t)
 \, = \, t^{w_i-w_j}M^{(i)}{}_{(j)}(G) \, .
\]
The second formula here is needed only for nonzero matrix entries, which have
nonnegative weight. Since the original $G$ satisfies the Bianchis, we obtain the 1-parameter family of Bianchis
\[
 \dd_MG_t^{(i)} \, = \, t^{w_i}\dd_MG^{(i)}
 \, = \, t^{w_i}P^{(i)}(G) \, = \, P^{(i)}(G_t) \, .
\]
For the flow equation, use the weighted trivialization identity to compute
\begin{align*}
 \dd_M\overline C_t^{(i)}
 +\sum_j\overline C_t^{(j)}M^{(i)}{}_{(j)}(G_t)
 &\, = \,t^{w_i-1}\left(w_i\dd_MC^{(i)}
                  +\sum_jw_jC^{(j)}M^{(i)}{}_{(j)}(G)\right)\\
 &\, = \,w_it^{w_i-1}\bigl(\dd_MC^{(i)}+R^{(i)}(C,G)\bigr)\\
 &\, = \,w_it^{w_i-1}G^{(i)}
 \, = \,\partial_tG_t^{(i)} \, .
\end{align*}
These are the two components of the full Bianchi identity on
$M\times I$ (cf. \eqref{WNC:eq:flow}). Pullback to either endpoint kills $\dd t$, and
$G_0=0$, $G_1=G$ by construction, hence $\sigma_W(C)$ is a null-concordance.

It remains to check that applying the specific recursion of
Proposition~\ref{WNC:prop:projection} recovers $C$ itself.
Denote its output along the interval by $C_{\mathrm{rec}}(t)$, to
distinguish it from the prescribed endpoint potential $C$.
We claim, by induction on weight, that
\[
 C_{\mathrm{rec}}^{(i)}(t) \, = \, t^{w_i}C^{(i)} \, .
\]
For this, it is sufficient to show that the correction $X^{(i)}$ in the recursion vanishes on these
candidates.

Suppose now that $C_{\mathrm{rec}}^{(j)}(t)=t^{w_j}C^{(j)}$
has been established at all lower weights. Every nonzero
$\partial_kM^{(i)}{}_{(j)}$ has weight $w_i-w_j-w_k$, so substitution
in \eqref{WNC:eq:X} gives
\begin{align*}
 X^{(i)}(t)
 &\, = \,\frac1{w_i}\sum_{j,k}w_j(-1)^{g^j}
 (t^{w_j}C^{(j)})(w_kt^{w_k-1}C^{(k)})
 t^{w_i-w_j-w_k}\partial_kM^{(i)}{}_{(j)}(G)\\
 &\, = \,\frac{t^{w_i-1}}{w_i}
 \sum_{j,k}w_jw_k(-1)^{g^j}C^{(j)}C^{(k)}     \partial_kM^{(i)}{}_{(j)}(G)
 \\
 & \, = \, \frac{t^{w_i-1}}{w_i}
\sum_{j,k}w_jw_k(-1)^{g^k}
C^{(k)}C^{(j)}
\partial_j\partial_kP^{(i)}(G)
\\
&\, = \, \frac{t^{w_i-1}}{w_i}
\sum_{j,k}w_jw_k
(-1)^{g^k+(g^j-1)(g^k-1)+g^jg^k}
C^{(j)}C^{(k)}
\partial_k\partial_jP^{(i)}(G)
\\
&\, = \, \frac{t^{w_i-1}}{w_i}
\sum_{j,k}w_jw_k(-1)^{g^j+1}
C^{(j)}C^{(k)}
\partial_k\partial_jP^{(i)}(G)
\, .
 \\
 \, &\, = \,
-X^{(i)} \, ,
\end{align*}
thus
$$
X^{(i)} \, = \,0 \, .
$$
Notice that at the lowest weights the correction sum is empty, giving the base
case by the same inductive calculation. The recursion and its initial value
therefore yield
\[
 C_{\mathrm{rec}}^{(i)}(t)
 \, = \,\int_0^t\overline C_u^{(i)}\,\dd u
 \, = \,\int_0^t w_i u^{w_i-1}C^{(i)}\,\dd u
 \, = \,t^{w_i}C^{(i)} \, ,
\]
which completes the induction. Evaluating at $t=1$ proves
\[
 P_W\bigl(\sigma_W(C)\bigr)^{(i)}
 \, = \,C_{\mathrm{rec}}^{(i)}(1)
 \, = \,C^{(i)} \, ,
\]
hence $P_W\sigma_W=\mathrm{id}$ and $P_W$ is surjective.
\end{proof}

\subsection*{Higher null-concordances and comparison to higher trivializations}
We wish to extend the above maps to an equivalence of $\infty$-groupoids, and so the first step is to define what constitutes higher concordances between null-concordances. But this is immediate from the definition of a null-concordance.
\begin{definition}[\bf Infinity groupoid of null-concordances]
\label{HigherNullConc}
The \textit{simplicial set of null-concordances} of a curvature $G$ on a manifold $M$ 
$$
\Null \, \equiv \, \Null^\infty(G)(M)
$$
has degree-$n$ set
\[
 (\Null)_n\, := \,
 \left\{\widehat{\mathcal{G}}\in\prod_i\Omega^{g^i}(M\times\Delta^n\times I_t)
 \;\middle|\;
 \dd_{M\times\Delta^n\times I_t }\widehat{\mathcal{G}}^{(i)} = P^{(i)}(\widehat{\mathcal{G}}), \qquad \widehat{\mathcal G}|_{t=0} \, = \,0,
 \qquad
 \widehat{\mathcal G}|_{t=1} \, = \,\operatorname{pr}_M^*G\right\} \, ,
\]
where we use smooth differential forms
up to the boundary, and the end-points equalities are as restrictions along $\mathrm{id}_{M\times \Delta^n}\times \iota_{t=\epsilon} \, : \, M\times \Delta^n \times \{\epsilon\} \hookrightarrow M\times \Delta^n\times I_t$, where $\epsilon=0,1$. The simplicial operators act via 
\[
\theta^*\widehat{\mathcal G}
\, := \,
(\mathrm{id}_M\times\theta_*\times\mathrm{id}_{I_t})^*
\widehat{\mathcal G} \, .
\]
for any non-decreasing $\theta:[m]\rightarrow[n]$.
\end{definition}

In particular, since $\Delta^1\cong I_s$,  $(\Null)_1$ consists of
concordances-of-null-concordances -- relative to the prescribed curvature endpoints (cf. \cite[Def.~2.46]{GSS24}\cite[Eqs.~(24-27)]{GSS26Survey}). Higher simplices then consist of coherent families of concordances between null-concordances and so on. The fact that this simplicial set is Kan is included in Step 4 of the proof of Thm. \ref{WNCH:thm:equivalence} below.

Now, notice that the construction of the weighted section map from Prop. \ref{WNC:prop:section} is natural under pullback in $M$. Replacing $M$ with  $M\times\Delta^n$ throughout makes the same
formulas available with higher simplex parameters so that $\dd_{M\times \Delta^n}$ 
includes all simplex directions, where differential forms are smooth up to the boundary when parameters
have boundary. It is thus natural to guess that both Prop.~\ref{WNC:prop:projection} and~\ref{WNC:prop:section} extend to a correspondence between higher concordances and higher trivializations.

However, upon a brief look at the formulas when replacing $M$ with $M\times \Delta^n$, one sees that these become maps between $n$-simplices of higher concordances and  $n$-simplices of \textit{total}-form trivializations $\Triv_\mathrm{total}$ on $M\times \Delta^n$ (Def. \ref{TotalTrivializations}), rather than $n$-simplices of trivializations $\Triv$ on $M$ (Def. \ref{TrivializationsGroupoid}). More precisely, we have the following extension result.
\begin{proposition}[\bf Higher weighted projection and section]
\label{WNCH:prop:extension}
Applying the weighted projection and section with base $M\times\Delta^n$ defines simplicial maps
\[
\Null
\ \underset{\sigma_W}{\overset{P_W}{\rightleftarrows}}\
\Triv_{\mathrm{total}},
\qquad
P_W\sigma_W\, =\,\mathrm{id}\, .
\]
\end{proposition}
\begin{proof}
For $\widehat{\mathcal G}\in\Null_n$, regard $\widehat{\mathcal G}$ as a $t$-null-concordance on the base $B_n=M\times\Delta^n$, with endpoints $0$ and $G_{B_n}=\operatorname{pr}_M^*G$. Proposition~\ref{WNC:prop:projection}, with the full differential $\dd_{B_n}$, gives
\[
\mathcal C\, :=\,P_{W,n}(\widehat{\mathcal G}),
\qquad
\dd_{B_n}\mathcal C+R(\mathcal C,G_{B_n})
\, =\,G_{B_n}\, ,
\]
so that $\mathcal C\in(\Triv_{\mathrm{total}})_n$. Note all simplex-direction components are retained -- only the distinguished concordance variable is integrated and then evaluated at $t=1$.

For a smooth map $f:\Delta^m\to\Delta^n$, put $f_M=\mathrm{id}_M\times f$. Pullback commutes with exterior differentiation, wedge products, and integration in $t$, and induction through the defining recursion therefore gives
\[
f_M^*P_{W,n}(\widehat{\mathcal G})
\, =\,
P_{W,m}\bigl((f_M\times\mathrm{id}_{I_t})^*\widehat{\mathcal G}\bigr)\, .
\]
Taking $f$ to be an affine coface or codegeneracy map $\theta_* : \Delta^m \rightarrow \Delta^n$ proves simplicial compatibility. In particular, every output face is the weighted projection of the corresponding input face.

Conversely, for $\mathcal C\in(\Triv_{\mathrm{total}})_n$, define
\[
\sigma_{W,n}(\mathcal C)^{(i)}
\, :=\,
t^{w_i}G_{B_n}^{(i)}
+w_it^{w_i-1}\dd t\,\mathcal C^{(i)}\, .
\]
Proposition~\ref{WNC:prop:section}, applied on $B_n$, proves that this lies in $\Null_n$ and that
\[
P_{W,n}\sigma_{W,n}\, =\,\mathrm{id}\, .
\]
The displayed section formula also commutes with parameter pullbacks, hence defines a simplicial map. Its endpoint restrictions are $0$ and $G_{B_n}$ as total forms.
\end{proof}

\smallskip\noindent\paragraph{\bf The case $n=1$.}
For exposition, and to connect to the trivialization model $\Triv$, we write out the case $n=1$. Writing $\Delta^1=I_s$, a concordance between $\widehat G_0$ and $\widehat G_1$ has boundary conditions
\[
\begin{aligned}
\widehat{\mathcal G}|_{s=0}&\, =\,\widehat G_0,
&\widehat{\mathcal G}|_{s=1}&\, =\,\widehat G_1\, ,\\
\widehat{\mathcal G}|_{t=0}&\, =\,0,
&\widehat{\mathcal G}|_{t=1}&\, =\,\operatorname{pr}_M^*G\, .
\end{aligned}
\]
These are equalities of total forms on $M\times I_t$ and $M\times I_s$, respectively, and express the fixed-endpoint condition of \cite[Def.~2.46]{GSS24}; see also \cite[Eqs.~(24)--(27)]{GSS26Survey}.
With parameter differentials on the left, write
\[
\widehat{\mathcal G}^{(i)}
\, =\,
F^{(i)}+\dd s\,A^{(i)}+\dd t\,B^{(i)}
+\dd t\wedge\dd s\,E^{(i)}\, ,
\]
where the coefficient forms on $M$ have degrees $g^i,g^i-1,g^i-1,g^i-2$. The $t$-endpoint conditions say
\[
F(s,0)\, =\,0,\qquad F(s,1)\, =\,G,
\qquad A(s,0)\, =\,A(s,1)\, =\,0\, ,
\]
and they place no additional endpoint restrictions on $B$ or $E$, since their terms contain $\dd t$. Restriction to $s=a$ gives $F(a,t)+\dd t\,B(a,t)=\widehat G_a$, for $a=0,1$.

Regrouping as
\[
\widehat{\mathcal G}
\, =\,(F+\dd s\,A)+\dd t\,(B+\dd s\,E)\, ,
\]
the projection on base $M\times I_s$ gives a total-form trivialization
\[
\mathcal C^{(i)}
\, =\,C^{(i)}(s)+\dd s\,\lambda_1^{(i)}(s),
\qquad |C^{(i)}|\, =\,g^i-1,\quad |\lambda_1^{(i)}|\, =\,g^i-2\, ,
\]
satisfying
\begin{equation}\label{WNCH:eq:output}
\begin{aligned}
(\dd_M+\dd_s)\mathcal C+R(\mathcal C,\operatorname{pr}_M^*G)
&\, =\,\operatorname{pr}_M^*G\, ,\\
C(a)&\, =\,P_{W,0}(\widehat G_a)\qquad(a=0,1)\, .
\end{aligned}
\end{equation}
The second line follows from pullback naturality. Expanding the first line of \eqref{WNCH:eq:output} yields
\[
\dd_M C+R(C,G)
+\dd s\bigl(\partial_sC-\dd_M\lambda_1+R(\lambda_1,G)\bigr)
\, =\,G\, ,
\]
and consequently
\[
\partial_s C\, =\,\DD_1^G\lambda_1,
\qquad
\Lambda_1\, :=\,\int_0^1\lambda_1(s)\,\dd s,
\qquad
C(1)-C(0)\, =\,\DD_1^G\Lambda_1\, .
\]

This last integration (cf. \eqref{Degree1Integration}) is the further passage to the gauge-parameter trivialization model (Def. \ref{TrivializationsGroupoid})
\[
I_1P_{W,1}(\widehat{\mathcal G})
\, =\,
\bigl(C(0),C(1),\Lambda_1\bigr)\, .
\]
Thus a concordance between null-concordances projects to a total-form $1$-simplex, and its integration gives the corresponding finite gauge transformation
$$
\Null_1 \xrightarrow{\quad P_{W,1}\quad } (\Triv_{\mathrm{total}})_1 \xrightarrow{\quad I_{1}\quad } \Triv_1 
$$
Conversely, the section sends a total-form $1$-simplex to a concordance between its endpoint section representatives, with the curvature endpoints fixed. Further composed with the Whitney map \eqref{Degree1Whitney}, it yields
$$
\Triv_1 \xrightarrow{\quad W_1\quad } (\Triv_\mathrm{total})_1 \xrightarrow{\quad \sigma_{W,1}} \Null_1
$$
The same composition holds at every simplicial degree, hence yielding the corresponding composites (Prop. \ref{IntegrationAndWhitneyRec}) of simplicial maps. 
\[
\Null
\ \underset{\sigma_W  W}{\overset{I P_W}{\rightleftarrows}}\
\Triv,
\qquad
I (P_W \sigma_W) W\, =\,\mathrm{id}\, .
\]
The goal is to show that these yield an equivalence of $\infty$-groupoids. This is a result of a composition of equivalences, as follows.

\paragraph{The intermediate equivalence.}
By
Proposition~\ref{WNCH:prop:extension}, the section and projection maps are simplicial and
$P_W\sigma_W=\mathrm{id}$. Thus constructing the weighted concordance of a trivialization and
projecting it back returns the original trivialization exactly. In the
other direction $\sigma_W P_W$, however, projection followed by the section generally changes
the concordance. Nevertheless, the resulting concordance is homotopic to the original one thus yielding an equivalence between the two $\infty$-groupoids.

\begin{theorem}[\bf Weighted trivializations and higher null-concordances]
\label{WNCH:thm:equivalence}
Under Assumption~\ref{ass:main}, let $G$ be a curvature on a smooth manifold $M$. 
\begin{enumerate}
\item The simplicial maps of Proposition~\ref{WNCH:prop:extension}
\[
 \Null
 \ \underset{\sigma_W}{\overset{P_W}{\rightleftarrows}}\
 \Triv_{\mathrm{total}}
\]
satisfy $P_W\sigma_W=\mathrm{id}$ and admit a simplicial homotopy
\[
 h:\Null\times\Delta[1]\longrightarrow\Null\, ,
 \qquad h|_0\, = \,\mathrm{id}\, ,
 \qquad h|_1\, = \,\sigma_WP_W\, ,
\]
which fixes $\operatorname{im}(\sigma_W)$ and preserves the curvature endpoints $0,G$.
\item Both simplicial sets are Kan, and hence these maps exhibit an equivalence of infinity groupoids
\[
\Null\,\simeq\,\Triv_{\mathrm{total}}\, .
\]
In particular, $\operatorname{im}(\sigma_W)$ is a strong simplicial deformation retract of $\Null$.
\end{enumerate}
\end{theorem}

\begin{proof}
Proposition~\ref{WNCH:prop:extension} gives the simplicial maps and the identity $P_W\sigma_W=\mathrm{id}$. It remains to construct a simplicial homotopy from $\mathrm{id}$ to $\sigma_WP_W$ which fixes the image of $\sigma_W$. All forms below are smooth up to the boundary, as usual.

\smallskip\noindent\emph{Strategy.}
We seek a curvature $\mathcal H$ on $M\times\Delta^n\times I_s\times I_t$ whose $s$-endpoint restrictions are $\widehat{\mathcal G}$ and $\sigma_WP_W(\widehat{\mathcal G})$, while its $t$-endpoint restrictions remain $0,G$. The point is that direct $s$-interpolation of the two curvatures $\widehat{\mathcal G}$ and $\sigma_WP_W(\widehat{\mathcal G})$ need not preserve their nonlinear Bianchi identities on the product with $I_s$. Instead, we retain the full \textit{potential family} constructed by the projection $P_W$ (cf. \ref{WNC:prop:projection}), interpolate it to the weighted family with the same endpoints, and reconstruct the curvature recursively. Using the total differential in this reconstruction supplies the $\dd s$-components required by the Bianchi identities on the product, and naturality then lets us apply the same passage from an interval family to a simplicial homotopy as in Theorem~\ref{WGPC:prop:comparison}. We carry out this construction and the Kan verification in four steps.

\smallskip\noindent
\emph{Step 1: reconstruction from potential families.}
Write $B=M\times\Delta^n$. We first show that from any smooth $t$-family of potentials on $B$, we may reconstruct a corresponding curvature satisfying the Bianchis on $B\times I_t$. Indeed, consider an \textit{arbitrary}  smooth family of would be ``potentials''
\[
 Q^{(i)}(t)\in\Omega^{g^i-1}(B)\, ,
\]
viewed as a form on $B\times I_t$ with no $\dd t$-component, and define $\mathcal K(Q)$ recursively in increasing weight by
\begin{equation}\label{WNCH:eq:reconstruction}
 \mathcal K(Q)^{(i)}
 \, := \,\dd_{B\times I_t}Q^{(i)}
       +R^{(i)}\bigl(Q,\mathcal K(Q)\bigr)\, .
\end{equation}
Every component on the right belonging to $R^{(i)}$ has strictly smaller weight than $w_i$, so this determines $\mathcal K(Q)$ uniquely.

Abbreviating $H=\mathcal K(Q)$ and $\dd=\dd_{B\times I_t}$ in the following, we inductively verify it satisfies the Bianchi identities. Suppressing the argument $H$ in the Bianchi polynomials and their derivative matrix, the lower-weight induction hypothesis and $\dd Q^{(j)}=H^{(j)}-R^{(j)}(Q,H)$ with $
R^{(i)}(Q,H)
:=
\frac1{w_i}\sum_jw_jQ^{(j)}M^{(i)}{}_{(j)}$ give
\begin{align*}
 \dd R^{(i)}(Q,H) &\,=\,
\frac1{w_i}\sum_jw_j
\left(
\dd Q^{(j)}M^{(i)}{}_{(j)}
+(-1)^{g^j-1}Q^{(j)}\dd M^{(i)}{}_{(j)}
\right).
 \\
 &\, = \,\frac1{w_i}\sum_jw_j
 \left((H^{(j)}-R^{(j)}(Q,H))M^{(i)}{}_{(j)}
       +(-1)^{g^j-1}Q^{(j)}\dd M^{(i)}{}_{(j)}\right)\\
 &\, = \,P^{(i)}(H)
 -\frac1{w_i}\sum_k w_k Q^{(k)}
 \left(\sum_jM^{(j)}{}_{(k)}M^{(i)}{}_{(j)}
       +(-1)^{g^k}\dd M^{(i)}{}_{(k)}\right)\\
 &\, = \,P^{(i)}(H)\, .
\end{align*}
Here the weighted Euler identity gives the first term in the third line, and the parentheses vanish by Lemma~\ref{lem:matrix}. All uses of the Bianchi identities in differentiating the matrix entries concern lower-weight components hence the induction is valid. Therefore 
$$
\dd H^{(i)}\, = \, \dd^2 Q^{(i)} + \dd R^{(i)}(Q,H) \, = \, P^{(i)}(H) \, ,
$$ 
where at the lowest weights this reduces to $H^{(i)}=\dd Q^{(i)}$ and $\dd H^{(i)}=0$. Note that the construction commutes with pullback in the base.

We now restrict to \textit{normalized} potential families. That is, suppose now that \(Q(0)=0\) and that \(Q(1)\) is a weighted trivialization of \(G_B\), i.e.,
\[
 Q(0)\, = \,0\, ,\qquad
 \dd_BQ(1)+R(Q(1),G_B)\, = \,G_B\, ,
 \qquad G_B:=\operatorname{pr}_M^*G\, .
\]
Then uniqueness of the same recursion \eqref{WNCH:eq:reconstruction} after endpoint restriction gives
\[
 \mathcal K(Q)|_{t=0}\, = \,0\, ,\qquad
 \mathcal K(Q)|_{t=1}\, = \,G_B\, ,
\]
thus 
$$
\mathcal K(Q)\in\Null_n \, .
$$

We next show that reconstruction of a curvature above is inverse to the full potential-family construction of Prop. \ref{WNC:prop:projection}, applied with base \(B\). Indeed, the latter assigns to $\widehat{\mathcal G}\in\Null_n$ a unique recursively constructed family $C_{\widehat{\mathcal G}}(t)$ such that
\[
 C_{\widehat{\mathcal G}}(0)\, = \,0\, ,\qquad
 \dd_{B\times I_t}C_{\widehat{\mathcal G}}
 +R(C_{\widehat{\mathcal G}},\widehat{\mathcal G})
 \, = \,\widehat{\mathcal G}\, ,\qquad
 C_{\widehat{\mathcal G}}(1)\, = \,P_W(\widehat{\mathcal G})\,  ,
\]
so that by \eqref{WNCH:eq:reconstruction}
$$
\mathcal K(C_{\widehat{\mathcal G}})=\widehat{\mathcal G}\, .$$ 
It remains to check that applying Prop. \ref{WNC:prop:projection} to the curvature \(\mathcal K(Q)\) recovers the original potential family \(Q(t)\). Writing $H=\mathcal{K}(Q)=H_0(t)+\dd t\,\overline C_H(t)$, the $\dd t$-coefficient of \eqref{WNCH:eq:reconstruction} is
\[
 \partial_tQ^{(i)}
 \, = \,\overline C_H^{(i)}
 +\frac1{w_i}\sum_{j,k}w_j(-1)^{g^j}
 Q^{(j)}\overline C_H^{(k)}
 \partial_kM^{(i)}{}_{(j)}(H_0)\, .
\]
But this is precisely the recursion \eqref{WNC:eq:recursiveC} defining the projection, with the same initial value $Q(0)=0$. Induction on weight shows that \(Q\) and \(C_H\) have the same derivative and the same zero initial value in every component, hence integration of their difference gives equality on the whole interval
$$
C_{\mathcal K(Q)}\, = \, C_H(t) \, =\, Q(t)\, .
$$

Reconstruction of curvatures and the full projected potential family are thus mutually inverse, naturally in $B$. The same argument applies to bases with corners, including $B\times I_s$.

\smallskip\noindent
\emph{Step 2: interpolation and its boundary restrictions.}
For any $\widehat{\mathcal G}\in\Null_n$, write
\[
 C(t):=C_{\widehat{\mathcal G}}(t)\, ,\qquad
 \mathcal C_*:=C(1)=P_W(\widehat{\mathcal G})\, ,
\]
and on $B\times I_s\times I_t$ define the interpolation potential family
\[
 Q_s^{(i)}(t)
 \, := \, (1-s)C^{(i)}(t)+s\,t^{w_i}\mathcal C_*^{(i)}\, .
\]
The point here is that, by Step 1, $C^{(i)}(t)$ is the potential family that reconstructs $\widehat{\mathcal G}$, while by Prop. \ref{WNC:prop:section}, $t^{w_i}\mathcal C_*^{(i)} = t^{w_i} P_W(\widehat{\mathcal{G}})$ is the family that reconstructs $\sigma_W P_W ({\widehat{\mathcal G}})$. Both families vanish at t = 0
and equal $\mathcal{C}_*$ at t = 1, so their $s$-interpolation preserves these potential endpoints.

We regard \(Q_s(t)\) as a smooth \(t\)-family of forms on the enlarged base \(B\times I_s\), and apply Step 1 with this base. That is, we reconstruct the total curvature of this interpolation using the differential in every direction
\[
 \mathcal H^{(i)}
 \, := \,\dd_{B\times I_s\times I_t}Q_s^{(i)}
       +R^{(i)}(Q_s,\mathcal H)\, .
\]
In particular, this is a total-form concordance, retaining $\dd s \partial_s Q_s^{(i)}$, and not merely a family of solutions for fixed $s$. Step 1 proves its Bianchi identities, and by uniqueness of reconstruction and Prop.~\ref{WNC:prop:section}, which identifies the projected family of $\sigma_W(\mathcal C_*)$ as $t^{w_i}\mathcal C_*^{(i)}$, its boundary can be seen to be
\[
 \begin{aligned}
 \mathcal H|_{s=0}&\, = \,\widehat{\mathcal G}\, ,
 &\mathcal H|_{s=1}&\, = \,\sigma_WP_W(\widehat{\mathcal G})\, ,\\
 \mathcal H|_{t=0}&\, = \,0\, ,
 &\mathcal H|_{t=1}&\, = \,G_B\, .
 \end{aligned}
\]
Pullbacks to the indicated products $B\times I_t$ and $B\times I_s$ on each row are understood. The second row holds as equalities of total forms on $B\times I_s$. In more detail, the four endpoint arguments are:
\begin{itemize}
\item At $s=0$, the potential family is $C(t)$, so uniqueness recovers $\widehat{\mathcal G}$.
\item At $s=1$, it is $t^{w_i}\mathcal C_*^{(i)}$, so Prop.~\ref{WNC:prop:section} and Step 1 recover $\sigma_W(\mathcal C_*)= \sigma_W P_W(\widehat{\mathcal G})$.
\item At $t=0$, the potential restricts to zero on $B\times I_s$, so the reconstructed curvature is zero.
\item At $t=1$, the potential restricts to the $s$-independent $\mathcal C_*=P_W(\widehat{\mathcal G})$. The pullback of $G_B$ solves that recursion on $B\times I_s$, and uniqueness excludes any extra $\dd s$-component.
\end{itemize}
If $\widehat{\mathcal G}=\sigma_W(\mathcal C_*)$, then $C^{(i)}(t)=t^{w_i}\mathcal C_*^{(i)}$, consequently $Q_s$ is independent of $s$
$$
Q_s^{(i)}(t)
 \, = \, (1-s)t^{w_i}\mathcal C_*^{(i)}+s\,t^{w_i}\mathcal C_*^{(i)} = t^{w_i}\mathcal C_*^{(i)}
$$
and $\mathcal H$ is simply the pullback of $\widehat{\mathcal G}$, so that the construction fixes $\operatorname{im}(\sigma_W)$. Moreover, Step 1 on the base $B\times I_s$ gives
\[
 P_W^{B\times I_s}(\mathcal H)\, = \,Q_s(1)\, = \,\mathcal C_*\, ,
\]
so the projection remains constant throughout the deformation.

\smallskip\noindent
\emph{Step 3: the simplicial homotopy.}
The preceding total-form construction is natural under simplex pullbacks along $\theta : [m] \rightarrow [n]$. Applying the barycentric-graph construction from Step 2 of Theorem~\ref{WGPC:prop:comparison}, with the additional $I_t$-coordinate carried along unchanged, gives a simplicial homotopy
\[
 h:\Null\times\Delta[1]\longrightarrow\Null\, ,
 \qquad h|_0\, = \,\mathrm{id}\, ,\qquad
 h|_1\, = \,\sigma_WP_W\, .
\]
The boundary and fixed-image properties established in Step 2 are preserved by these pullbacks.

\smallskip\noindent
\emph{Step 4: the Kan property and conclusion.}
The total-form model $\Triv_\mathrm{total}$ is Kan by Lem.~\ref{prop:WeightedTrivKan}. To prove this independently for $\Null$, note that Step 1 identifies it simplicially with the space of smooth potential families $Q(t)$ satisfying
\[
 Q(0)\, = \,0\, ,\qquad
 \dd_{B}Q(1)+R(Q(1),G_B)\, = \,G_B\, ,
\]
which are affine linear conditions on $Q$. Thus , if a vertex $C_0\in(\Triv_{\mathrm{total}})_0$ exists, the families
\[
 Q_{0,n}^{(i)}(t)\, := \,t^{w_i}\operatorname{pr}_M^*C_0^{(i)}
\]
give
compatible origins (in the potential-family model) for $\Null_n$ in every simplicial degree. Subtraction identifies the family model with a simplicial vector space, which is Kan, while if no vertex exists, both simplicial sets are empty by the maps $P_W,\sigma_W$, and are Kan vacuously.

Thus the displayed homotopy and $P_W\sigma_W=\mathrm{id}$ exhibit homotopy inverse maps of Kan complexes. Since the homotopy fixes the image of $\sigma_W$, it gives the asserted strong simplicial deformation retraction.
\end{proof}

\begin{corollary}[\bf Gauge parameters and homotopy groups of null-concordances]
\label{WNCH:cor:gaugehomotopy}
Under the hypotheses of Theorem~\ref{WNCH:thm:equivalence}:
\begin{enumerate}
\item
The maps
\[
 I P_W:\Null\longrightarrow\Triv,
 \qquad
 \sigma_W W:\Triv\longrightarrow\Null
\]
are homotopy inverses. Hence
\[
 \Null \,\simeq\, \Triv \, ,
\]
and in particular, $G$ admits a null-concordance if and only if
it admits a global weighted trivialization.

\item
Suppose these equivalent conditions hold. Then the set of connected
components 
$$
\pi_0\Null
$$ is naturally a torsor for
\[
 H^0(L^\bullet)
 \, = \,
 \frac{\ker\DD_0^G}{\operatorname{im}\DD_1^G},
\]
where $L^\bullet$ is the weighted gauge complex of
Definition~\ref{WGPC:def:gaugecomplex}.

Choosing a null-concordance $\widehat G_*$ and putting
$C_*:=P_W(\widehat G_*)$ identifies this torsor with
$H^0(L^\bullet)$ by
\[
 [\widehat G]
 \,\longmapsto\,
 [P_W(\widehat G)-C_*] \, .
\]
Thus two null-concordances lie in the same component precisely
when their projected potentials differ by a
$\DD_1^G$-exact term -- namely a finite gauge transformation.

\item
For any basepoint $\widehat G_*\in\Null_0$ and every $r\geq1$,
\[
 \pi_r(\Null,\widehat G_*)
 \,\cong\,
 H^{-r}(L^\bullet)
 \, = \,
 \frac{\ker\DD_r^G}{\operatorname{im}\DD_{r+1}^G} \, .
\]
Here $\DD_r^G$ acts on
$\mathcal A_r(M)=\prod_i\Omega^{g^i-1-r}(M)$,
with negative-degree forms understood to vanish.

In the gauge-parameter model, these classes are represented
by parameters $\Lambda_r$ satisfying
$\DD_r^G\Lambda_r=0$, modulo shifts by
$\DD_{r+1}^G\Lambda_{r+1}$.
Their group operation is induced by addition, and so,
 in particular, $\pi_1(\Null,\widehat G_*)$ is abelian.
\end{enumerate}
\end{corollary}

\begin{proof}
For the first assertion, combine
Theorem~\ref{WNCH:thm:equivalence} with
Theorem~\ref{WGPC:prop:comparison}.
The identities $P_W\sigma_W=\mathrm{id}$ and
$IW=\mathrm{id}$ give
\[
 (IP_W)(\sigma_W W) \, = \, \mathrm{id}_{\Triv},
\]
while the other composite is homotopic to the identity
by the two comparison homotopies.

For the remaining assertions, choose a trivialization $C_*=P_W(\widehat G_*)$.
By Remark~\ref{WGPC:rem:doldkan}, translation by $C_*$
identifies $\Triv$ with the Dold--Kan simplicial vector
space of the gauge complex, with the indicated degreewise
sign convention. By Kan’s homotopy–homology identification \cite[Cor.~8.3]{Kan1958}, its \(r\)-th homotopy group is \(H^{-r}(L^\bullet)\). Together with Thm \ref{WNCH:thm:equivalence} and the comparison theorem, this yields the claimed identification.

Transporting these identifications along $IP_W$ gives the
stated formulas for $\Null$, while without choosing an origin,
the connected components form the stated torsor.
\end{proof}


\newpage

\begin{thebibliography}{10}

\bibitem[Alb20]{Albert}
B.~I.~Albert,
\emph{\color{darkblue}The Dupont Homotopy Formula and Stellar Subdivision},
(2020).
\href{https://arxiv.org/abs/1902.00627v2}{[arXiv:1902.00627]}.

\bibitem[BS17]{BandieraSchaetz}
R.~Bandiera and F.~Sch\"atz,
\emph{\color{darkblue}How to discretize the differential forms on the interval},
Higher Structures \textbf{1}(1) (2017), 56--86.
\href{https://doi.org/10.21136/HS.2017.03}{doi:10.21136/HS.2017.03}.

\bibitem[CL94]{CL94}
A. Candiello and K. Lechner, 
{\it \color{darkblue} Duality in supergravity theories}, Nucl. Phys. B {\bf 412}  (1994), 479-501, \newline [\href{https://doi.org/10.1016/0550-3213(94)90389-1}{\tt doi:10.1016/0550-3213(94)90389-1}],
[\href{https://arxiv.org/abs/hep-th/9309143}{\tt arXiv:hep-th/9309143}].

\bibitem[CS24]{CiriciSaleh}
J.~Cirici and B.~Saleh,
\emph{\color{darkblue}Weight Decompositions on Algebraic Models for Mapping Spaces and
Homotopy Automorphisms},
Mediterranean Journal of Mathematics \textbf{21} (2024), article 156.
\href{https://doi.org/10.1007/s00009-024-02700-6}{doi:10.1007/s00009-024-02700-6};
\href{https://arxiv.org/abs/2308.12821}{[arXiv:2308.12821]}.

\bibitem[Del86]{Deligne1986}
P.~Deligne, \emph{\color{darkblue}Letter to J.~Millson on controlling differential graded
Lie algebras} (1986), unpublished letter, scan hosted by J.~Millson.
\url{https://www.math.umd.edu/~millson/papers/deligne.pdf}.

\bibitem[Dup78]{Dupont78}
J.~L.~Dupont,
\emph{\color{darkblue}Curvature and Characteristic Classes},
Lecture Notes in Mathematics \textbf{640}, Springer-Verlag (1978).
\href{https://doi.org/10.1007/BFb0065364}{doi:10.1007/BFb0065364}.

\bibitem[FSS15]{FSS15-M5WZW}
D. Fiorenza, H. Sati, and U. Schreiber,
{\it \color{darkblue} The WZW term of the M5-brane and differential cohomotopy}, 
J. Math. Phys. {\bf 56} (2015)  102301,
[\href{https://doi.org/10.1063/1.4932618}{\tt doi:10.1063/1.4932618}],
[\href{https://arxiv.org/abs/1506.07557}{\tt arXiv:1506.07557}].

\bibitem[FSS17]{FSS17}
D. Fiorenza, H. Sati, and U. Schreiber, {\it \color{darkblue} Rational sphere valued supercocycles in M-theory and type IIA string theory}, 
J. Geom. Phys.
{\bf 114} (2017), 91-108, 
[\href{http://dx.doi.org/10.1016/j.geomphys.2016.11.024}{\tt doi:10.1016/j.geomphys.2016.11.024}],
[\href{https://arxiv.org/abs/1606.03206}{\tt arXiv:1606.03206}].

\bibitem[FSS23]{FSS23Character}
D.~Fiorenza, H.~Sati, and U.~Schreiber,
\emph{\color{darkblue}The Character Map in Non-Abelian Cohomology:
Twisted, Differential, and Generalized}, World Scientific (2023).
\href{https://doi.org/10.1142/13422}{doi:10.1142/13422};
\href{https://arxiv.org/abs/2009.11909}{[arXiv:2009.11909]}.


\bibitem[Get09]{Getzler}
E.~Getzler,
\emph{\color{darkblue}Lie theory for nilpotent $L_\infty$-algebras},
Annals of Mathematics \textbf{170} (2009), 271--301.\newline 
\href{https://doi.org/10.4007/annals.2009.170.271}{doi:10.4007/annals.2009.170.271};
\href{https://arxiv.org/abs/math/0404003}{[arXiv:0404003]}.

\bibitem[Gi25]{Gi25}
G. Giotopoulos, {\it\color{darkblue}Sheaf Topos Theory: A powerful setting for Lagrangian Field Theory}, 
[\href{https://arxiv.org/abs/2504.08095}{\tt arXiv:2504.08095}].




\bibitem[GSS24]{GSS24}
G. Giotopoulos, H. Sati, and U. Schreiber, 
{\it\color{darkblue} Flux Quantization on 11d Superspace},
J. High Energy Phys. {\bf 2024} 82 (2024)
[\href{https://doi.org/10.1007/JHEP07(2024)082}{\tt doi:10.1007/JHEP07(2024)082}],
[\href{https://arxiv.org/abs/2403.16456}{\tt arXiv:2403.16456}].

\bibitem[GSS24b]{GSS-M5Brane}
G. Giotopoulos, H. Sati, and U. Schreiber, 
{\it \color{darkblue}Flux Quantization on M5-Branes},
J. High Energy Physics {\bf 2024} (2024) 140,
[\href{https://doi.org/10.1007/JHEP10(2024)140}{\tt doi:10.1007/JHEP10(2024)140}],
[\href{https://arxiv.org/abs/2406.11304}{\tt arXiv:2406.11304}].

\bibitem[GS26]{GSIIA}
G. Giotopoulos and H. Sati
{\it\color{darkblue} Flux Quantization on 10D Type IIA Superspace via Cyclification from 11D} 
[\href{https://arxiv.org/abs/2606.16857}{\tt arXiv:2606.16857}].

\bibitem[GSS26]{GSS26Survey}
G. Giotopoulos, H. Sati, and U. Schreiber, 
{\it \color{darkblue}Higher Superspace Supergravity and its IR Completions} 
[\href{https://arxiv.org/abs/2607.20312}{\tt arXiv:2607.20312}].

\bibitem[GSS26b]{GSS24-TDuality}
G. Giotopoulos, H. Sati, and U. Schreiber, 
{\it\color{darkblue}Super-$\mathrm{Lie}_{\infty}$ T-Duality and M-Theory},
Rev. Math. Phys. {\bf  38} (2026) 2550018,
[\href{https://doi.org/10.1142/S0129055X25500187}{\tt 
doi:10.1142/S0129055X25500187}], 
[\href{https://arxiv.org/abs/2411.10260}{\tt 	arXiv:2411.10260}].

\bibitem[Hen08]{Henriques}
A.~Henriques,
\emph{\color{darkblue}Integrating $L_\infty$-algebras},
Compositio Mathematica \textbf{144} (2008), 1017--1045.\newline
\href{https://arxiv.org/abs/math/0603563}{[arXiv:0603563]}.

\bibitem[Hin97]{Hinich1997}
V.~Hinich, \emph{\color{darkblue}Descent of Deligne groupoids},
International Mathematics Research Notices \textbf{1997}, no.~5, 223--239.
\href{https://doi.org/10.1155/S1073792897000160}
{doi:10.1155/S1073792897000160};
\href{https://arxiv.org/abs/alg-geom/9606010}{[arXiv:9606010]}.



\bibitem[Kan]{Kan1958}
D.~M.~Kan,
\emph{\color{darkblue}Functors involving c.s.s. complexes},
Transactions of the American Mathematical Society
\textbf{87} (1958), no.~2, 330--346.
\href{https://doi.org/10.1090/S0002-9947-1958-0131873-8}
{doi:10.1090/S0002-9947-1958-0131873-8}.

\bibitem[Ker]{Kerodon}
The Kerodon Project.
\emph{Kerodon}.
Proposition~1.2.5.9, ``Kan Complexes''.
\href{https://kerodon.net/tag/00MG}{Tag~00MG}
(accessed 27 October 2026).

\bibitem[Lun18]{Lunardon}
L.~Lunardon, \emph{\color{darkblue} Some remarks on Dupont contraction},
Rendiconti di Matematica e delle sue Applicazioni (7)
\textbf{39} (2018), pp.~79--96.
\href{https://arxiv.org/abs/1807.02517}{[arXiv:1807.02517]}.

\bibitem[Sa13]{Sati13}
H. Sati, 
{\it\color{darkblue} Framed M-branes, corners, and topological invariants}, 
J. Math. Phys. {\bf 59} (2018) 062304, \newline 
[\href{https://doi.org/10.1063/1.5007185}{\tt doi:10.1063/1.5007185}], 
[\href{https://arxiv.org/abs/1310.1060}{\tt arXiv:1310.1060}].

\bibitem[Sc13]{dcct}
U. Schreiber,
{\it \color{darkblue} Differential cohomology in a cohesive infinity-topos},
[\href{https://arxiv.org/abs/1310.7930}{\tt arXiv:1310.7930}].

\bibitem[Sc24]{Schreiber24}
U. Schreiber, 
{\it\color{darkblue}Higher Topos Theory in Physics},
{\it Encyclopedia of Mathematical Physics}, 2nd ed., Elsevier (2024),
62-76, 
[\href{https://doi.org/10.1016/B978-0-323-95703-8.00210-X}{\tt 
doi:10.1016/B978-0-323-95703-8.00210-X}], 
[\href{https://arxiv.org/abs/2311.11026}{\tt arXiv:2311.11026}].

\bibitem[Sta]{Stacks}
The Stacks Project Authors.
\emph{The Stacks Project}.
Section~14.26, ``Homotopies''.
\href{https://stacks.math.columbia.edu/tag/019J}{Tag~019J}
(accessed 27 October 2026).

\bibitem[SS25]{SS25Flux}
H.~Sati and U.~Schreiber,
\emph{\color{darkblue}Flux Quantization}, Encyclopedia of Mathematical Physics,
2nd ed., vol.~4 (2025), 281--324.
\href{https://arxiv.org/abs/2402.18473}{[arXiv:2402.18473]}.


\bibitem[SV23]{SV1}
H. Sati  and  A. Voronov, 
{\it \color{darkblue}Mysterious Triality and Rational Homotopy Theory}, 
Commun. Math. Phys. {\bf 400} (2023), 1915–1960,
[\href{https://doi.org/10.1007/s00220-023-04643-7}{\tt doi:10.1007/s00220-023-04643-7}],
[\href{https://arxiv.org/abs/2111.14810}{\tt  	arXiv:2111.14810}]. 

\bibitem[SV25]{SV-loop}	
H. Sati and A. Voronov,
{\it\color{darkblue}Mysterious Triality and the Exceptional Symmetry of Loop Spaces},
	Lett. Math. Phys. {\bf 115} (2025) 98,
[\href{https://doi.org/10.1007/s11005-025-01977-2}{\tt 
doi:10.1007/s11005-025-01977-2}], 
[\href{https://arxiv.org/abs/2408.13337}{\tt arXiv:2408.13337}].


\bibitem[Sul77]{Sullivan1977}
D.~Sullivan,
\emph{\color{darkblue} Infinitesimal computations in topology},
Publ. Math. IH\'ES \textbf{47} (1977),269--331.\newline 
\href{https://doi.org/10.1007/BF02684341}{doi:10.1007/BF02684341}
\end{thebibliography}
\end{document}